\documentclass[prb,aps,twocolumn,superscriptaddress,showpacs]{revtex4}
\usepackage{graphicx}
\usepackage{amsmath}
\usepackage{wasysym}
\usepackage{color}
\usepackage{bm}

\newcommand{\rmi}{{\rm i}}
\newcommand{\veck}{{\bf k}}
\newcommand{\veci}{{\bf i}}
\newcommand{\vecj}{{\bf j}}

\newcommand{\vecQ}{{\bf Q}}

\newcommand{\delpi}{{\Delta_{\bf \vecQ}}}
\newcommand{\npi}{{n_{\bf \vecQ}}}
\newcommand{\opdelpi}{{\hat{\Delta}_{\bf \vecQ}}}
\newcommand{\opnpi}{{\hat{n}_{\bf \vecQ}}}
\begin{document} 
\title{Charge and pairing dynamics in the attractive Hubbard model: mode coupling and the validity of linear-response theory} 
\author{J\"org B\"unemann}
 \affiliation{Institut f\"ur Physik, BTU Cottbus-Senftenberg, P.O.\ Box 101344, 03013 Cottbus, 
Germany}
\affiliation{Fakult\"at Physik,
Technische Universit\"at Dortmund, 44227 Dortmund, Germany}
 \author{G\"otz Seibold}
 \affiliation{Institut f\"ur Physik, BTU Cottbus-Senftenberg, P.O.\ Box 101344, 03013 Cottbus, Germany}
 
\date{\today}
 
 
\begin{abstract} 
  Pump-probe experiments have turned out as a powerful tool in
  order to study the dynamics of competing orders in a large
  variety of materials. The corresponding analysis of the data
  often relies on standard linear-response theory generalized to
  non-equilibrium situations. Here we examine the validity of
  such an approach within the attractive Hubbard model for which
  the dynamics
  of pairing and charge-density wave orders is
  computed using the time-dependent Hartree-Fock approximation (TDHF).
  Our calculations reveal that the `linear-response
  assumption' is justified for small to moderate non-equilibrium
  situations (i.e., pump pulses) when the symmetry of the pump-induced state
  differs from that of the external field. This is the case, when we
  consider the pairing response in a charge-ordered state or the
  charge-order response in a superconducting state. The situation is
 very different when the non-equilibrium state  and the external probe field 
have the same symmetry. In this case, we observe significant changes of the 
response in magnitude but also due to mode coupling when moving away
from an equilibrium state, indicating the
failure of the linear-response assumption.
\end{abstract} 
 
\maketitle 
\section{Introduction}\label{sec_true_intro}
In many experiments that are carried out in solid-state physics, 
 one measures so called `response functions'. Such a function provides 
 information on the {\it linear} response of a given observable to
 a small time (or frequency)  dependent external perturbation.
 When a system is in its ground state (or, at finite temperature, in
 thermal equilibrium) 
 the well-known 'Kubo formula'~\cite{kubo1957,kubo1959}
identifies the response functions  as retarded 
two-particle Greens functions.~\cite{mahan2005} Important examples are  
 the magnetic or the charge susceptibilities as well as the optical 
conductivity. 

 In recent years, the development of ultrafast laser sources  made it possible
 to measure response functions not only in equilibrium. Such measurements 
  are usually denoted as `pump-and-probe experiments' because a 
 large pump pulse first drives the system out of equilibrium 
 before a small probe puls measures the usual response function.
 This kind of technique has been successfully applied to investigate
 the dynamics of electronic and phononic processes in high-T$_c$
 superconductors \cite{pupro1,pupro2,pupro3,pupro4,pupro5,pupro6,pupro7,pupro8} in order to elucidate the 'glue' for the Cooper pair binding.
 This can be achieved e.g. by a pump pulse through impulsive stimulated
 Raman scattering which induces an out-of-equilibrium condensate for which
 coupled excitations can be measured by a successive optical probe \cite{mansart13,lorenzana13}.
 Moreover, pump and probe methods have been used to study the out-of equilibrium
 dynamics of competing order parameter in correlated
 systems as e.g. the dynamics of
 spin and charge orders in nickelates \cite{chuang13,lee17} or the interplay of 
charge-density
 wave and superconducting orders in high-T$_c$ cuprates \cite{foerst14,foerst142} 
 (for a review see Refs.~[\onlinecite{pumpprobe1},\onlinecite{manko16}] and references therein).  

 It is obvious that the calculation of a non-equilibrium response function 
 is even more challenging than that of its equilibrium counterpart.
In this regard, different schemes have been employed to generalize the
 Kubo formula to out-of-equilibrium situations \cite{filippis12,rossini14,lenarcic14}
 which have been critically analyzed in the context of the optical
 conductivity. \cite{takami16}

 In general, the dynamics of a quantum system can be obtained
 from non-equilibrium Green's function (NEGF) techniques
 which requires the solution of the so-called Keldysh-Kadanoff-Baym equations.
 \cite{kadbaym62,keld64} For interacting systems these
 are usually decoupled within a conserving approximation.
 For weak to moderate interactions the lowest order corresponds
 to the time-dependent Hartree-Fock approximation (TDHF) which will be
 employed in the present paper. We note that in case of
 strongly correlated systems also the dynamical mean-field theory
  \cite{aoki14} and the Gutzwiller approximation \cite{schiro2010,schiro2011,buenemann2013a}
 have been generalized to the description of time-dependent phenomena. 
 On the other hand, in symmetry-broken systems the
 order parameter dynamics can be phenomenologically described through
 time-dependent Ginzburg-Landau theory (see e.g. Ref.~[\onlinecite{pumpprobe1}])
 which among others has been successfully applied to the description of
 spin and charge order dynamics in  nickelates. \cite{chuang13,kung13,lee17}

 In this work, it is our aim to analyze the dynamics of competing orders
 in a given microscopic model which, as mentioned above,  we accomplish
 within the TDHF. \cite{ring1980,blaizot1986}
 This method  can be used for the study of time development
 near and away from equilibrium. It therefore allows us the unbiased  
investigation 
 of pump-and-probe situations without any linear-response assumption. 
Moreover,
 in the small-amplitude limit the
 TDHF reduces to the well-known `random-phase approximation' \cite{ring1980,blaizot1986}  which corresponds to linear-response theory so that
 our approach allows for exploring the validity of the Kubo formula
 in out-of equilibrium situations of competing orders.
 The major drawback of the Hartree-Fock approximation 
is its use of single-particle wave functions which can lead 
to flawed results already for the ground-state properties when
correlation effects become important.

Our investigations will focus on the attractive Hubbard model
which is one of the simplest
  systems that shows non-trivial symmetry-broken phases already within 
  the Hartree-Fock approximation. For weak on-site attraction
  and at zero temperature
  the dominating instability is that to a standard BCS superconductor
  with isotropic s-wave order parameter. For a bipartite lattice at
  half-filling the model has $SO(3)$ symmetry and the SC state
  is degenerate with a commensurate charge-density wave.
  Without additional long-range interactions and away from half-filling,
 the SC phase constitutes
  the ground state, i.e., it is always more stable than a charge ordered
 state. \cite{micnas90}

  The SC order parameter dynamics of related BCS-type models has
  been in the focus of numerous previous studies \cite{barankov04,papenkort07,papenkort08,krull14} and has also been investigated for 
  multiband superconductivity. \cite{akbari13}
  In the linear-response limit \cite{volkov73} a small perturbation of the
  SC order parameter $\Delta$ excites amplitude modes with an energy 
corresponding
  to the SC gap $2\Delta$ and which are damped due to their admixture
  with Bogoljubov quasiparticle excitations. As a consequence the order
  parameter relaxation towards a constant shows an oscillatory behavior with frequency $\Omega=2\Delta$ and an amplitude which decays $\sim 1/\sqrt{t}$. Later it has been shown \cite{altshuler06} that this dynamics is also obeyed beyond the linear
  response regime when the non-equilibrium state is in the same class as
  the ground state, e.g., when the non-equilibrium state is generated from a paired ground state by a sudden change of the pairing strength. In the other case, 
i.e., when non-equilibrium and
  ground state are topologically different, persistent oscillations of the
  order parameter occur. This can be achieved, e.g., by an initial normal state
  while the ground state is a superconductor.

  In the present paper, we investigate the dynamics of competing orders,
  namely charge density wave (CDW) order and SC order, in the context of the
  validity of linear-response theory in a non-equilibrium situation.
  Thus we will also deal with a scenario where a ground and non-equilibrium
  state have different symmetries, namely SC and CDW or vice versa.
  However, since without additional interaction the ground state of the
  attractive Hubbard model at weak coupling is a BCS superconductor we
  supplement the model with a staggered charge order field, arising, 
e.g., 
  from a lattice distortion, in order
  to realize three different ground state symmetries, i) 
a pure SC state, ii)
 a charge-ordered state, or iii) a state with both orders present. 
 The stability and proximity of these phases makes the   
attractive Hubbard model an ideal playing field for the 
 study of non-equilibrium response functions because we can combine 
 each of the three possible symmetries of the pump-pulse induced 
 initial state with the two relevant symmetries of a probe pulse.
  Note that the attractive Hubbard model with both SC and CDW orders
 has recently also been investigated in the context of
the visibility of the amplitude (Higgs) mode within 
 linear (Raman) response. \cite{cea142} 
 
Our work is organized as follows: In Sec.~\ref{theory} we discuss our model, 
 the details about its treatment within TDHF, and our way of simulating 
pump and 
probe experiments. 
 The ground-state properties of our model are discussed in Sec.~\ref{gs}. 
In Sec.~\ref{ooed}, on out-of-equilibrium dynamics, we show in detail
the results for quantum-quench dynamics as well as pump-and-probe simulations.  
We close our presentation with concluding remarks in Sec.~\ref{conclusions}. Details on 
 our numerical minimization and the results for two tutorial toy models 
are  deferred to three appendices. 
   
\section{Model and Method}\label{theory}
In this chapter, we will present the theoretical background of our study. 
 First, in Sec.~\ref{ham}, we introduce our model and derive its Hartree-Fock
 energy functional. Second, in Sec.~\ref{theory2}, the TDHF equations for 
 our model are derived. Third, in Sec.~\ref{qyx},  we explain how we will
 simulate pump and probe experiments.
\subsection{Hamiltonian and ground-state energy functional}\label{ham}
We consider the attractive (`negative $U$') Hubbard model defined by
\begin{eqnarray}\label{sdf}
\hat{H}_{\rm H}&=&\hat{H}_0 - U\sum_{\veci}  \hat{n}_{\veci,\uparrow}
\hat{n}_{\veci,\downarrow}\;\;(U\geq 0)\;,\\\label{sdf2}
\hat{H}_0&=&\sum_{\veci,\vecj,\sigma}t_{\veci,\vecj}\hat{c}_{\veci,\sigma}^{\dagger}
\hat{c}_{\vecj,\sigma}^{\phantom{\dagger}}= \sum_{\veck,\sigma}
\varepsilon_{\veck}
\hat{c}_{\veck,\sigma}^{\dagger}
\hat{c}_{\veck,\sigma}^{\phantom{\dagger}}
\end{eqnarray}
where $\veci,\vecj $ are lattice-site vectors, $\sigma$ the spin index, 
 $\hat{n}_{\veci,\sigma}\equiv \hat{c}_{\veci,\sigma}^{\dagger} \hat{c}_{\veci,\sigma}^{\phantom{\dagger}} $, and $\veck$ a wave vector in the first Brillouin zone.  
For simplicity, we assume a bipartite lattice with sublattices  $A$/$B$ 
 that can be defined with a nesting vector $\vecQ$  via
\begin{equation}
e^{\rmi \vecQ\cdot \veci}= \left \{ \begin{array}{c}
+1\;\;\;\;\;\;{\rm if}\;\; \veci \in A\\
-1\;\;\;\;\;\;{\rm if}\;\;\veci \in B\\
\end{array}
\right. \;\;.
\end{equation}
We further assume that the hopping parameters $t_{\veci,\vecj}$ are non-zero 
only when $\veci$ and $\vecj$ belong to different sublattices. This leads to 
  \begin{equation}\label{disp}
\varepsilon_{\veck+\vecQ}=-\varepsilon_{\veck}
\end{equation}
for the 
dispersion relation in~(\ref{sdf2}). 

In the following, we want to study states
  which may include local pairing as well as charge order. On a 
 mean-field level, i.e., evaluated with a single-particle product 
wave function, the expectation value of the (local) two-particle 
 interaction  in~(\ref{sdf}) then has the form
\begin{equation}
\langle \hat{n}_{\veci,\uparrow}
\hat{n}_{\veci,\downarrow} \rangle =
\langle \hat{n}_{\veci,\uparrow}
\rangle\langle 
\hat{n}_{\veci,\downarrow} \rangle
+\langle 
\hat{c}_{\veci,\uparrow}^{\dagger} \hat{c}_{\veci,\downarrow}^{\dagger} 
\rangle
\langle 
\hat{c}_{\veci,\downarrow}^{} \hat{c}_{\veci,\downarrow}^{} 
\rangle
\end{equation}
where we impose the charge- and pair-density fields as
\begin{eqnarray}
\langle \hat{c}_{\veci,\uparrow}^{\dagger} \hat{c}_{\veci,\downarrow}^{\dagger} \rangle 
&=&\Delta_0+e^{\rmi \vecQ \cdot \veci}\delpi \;,\\
\langle \hat{n}_{\veci,\sigma}
\rangle&=&n_0+e^{\rmi \vecQ \cdot \veci}\npi\;.
\end{eqnarray}
Here, $\Delta_0$, $n_0$, $\npi$, and $\delpi$ are 
lattice-site independent numbers. 

In real materials, a charge order can be stabilized by a static distortion 
 of the lattice. To simulate this effect we allow for a external
 `charge-order field'  
\begin{equation}\label{rta}
\hat{H}_{\rm co}\equiv \frac{\alpha_{\vecQ}}{2}\sum_{\veci,\sigma}
e^{\rmi \vecQ\cdot \veci}\hat{n}_{\veci,\sigma}
\end{equation}
and study in the following the Hamiltonian
\begin{equation}
\hat{H}\equiv  \hat{H}_{\rm H}+\hat{H}_{\rm co}\;.
\end{equation}

In a superconducting phase the  total particle number is not conserved 
 but its expectation value has to be fixed by means of 
a chemical potential $\mu$.  
 Hence we work with $\hat{K}\equiv \hat{H}-\mu \hat{N}$ instead of $\hat{H}$. 
 The expectation value of $\hat{K}$  is given as
\begin{eqnarray}\label{qwe}
\frac{\langle \hat{K}  \rangle}{L}&=&
\frac{1}{L}
\sum_{\veck,\sigma}
(\varepsilon_{\veck}-\mu)
\langle 
\hat{c}_{\veck,\sigma}^{\dagger}
\hat{c}_{\veck,\sigma}^{\phantom{\dagger}} \rangle\\\nonumber
&&-U\Big(
\big(\npi)^2+|\Delta_0|^2+|\delpi|^2
\Big)+\alpha_{\vecQ}\npi\;.
\end{eqnarray}
Note that in this expression
 we have dropped the constant energy shift $Un_0^2$ on the right hand side. The 
 order parameters in~(\ref{qwe}) can be calculated in momentum space with
\begin{eqnarray}\label{yxc}
\Delta_0&=&\langle \hat{\Delta}_0\rangle\;\;,\;\;
\hat{\Delta}_0\equiv\frac{1}{L}\sum_{\veck} \hat{c}_{\veck,\uparrow}^{\dagger}
 \hat{c}_{-\veck,\downarrow}^{\dagger}
 \;, \\\label{yxcb}
\delpi&=&\langle\opdelpi\rangle\;\;,\;\opdelpi\equiv \frac{1}{L}\sum_{\veck} \hat{c}_{\veck,\uparrow}^{\dagger}
 \hat{c}_{-\veck-\vecQ,\downarrow}^{\dagger}\;,
 \\ \label{yxc2}
   \npi  &=&\langle \opnpi\rangle \;\;,\;\; \opnpi\equiv \frac{1}{2 L}\sum_{\veck,\sigma}
 \hat{c}_{\veck,\sigma}^{\dagger}
 \hat{c}_{\veck+\vecQ,\sigma}^{\phantom{\dagger}}  \;.
\end{eqnarray}
With the nesting vector $\vecQ$, we may split the Brillouin zone 
$\mathcal B$ into 
two parts ${\mathcal B}={\mathcal B}_0\cup {\mathcal B}_{\vecQ}$ 
such that for each $\veck$ we have either $\veck\in {\mathcal B}_0$
or $\veck\in {\mathcal B}_{\vecQ}$, $\veck-\vecQ\in {\mathcal B}_{0}$. 
For convenience, ${\mathcal B}_0$ is chosen such that, 
with $\veck \in {\mathcal B}_0$ it is also  $-\veck \in {\mathcal B}_0$.
   To get rid of the anomalous expectation values  
 in~(\ref{yxc}),(\ref{yxcb}) we introduce the following canonical transformation
\begin{eqnarray}
\hat{d}_{\veck,1}^{\dagger}&=&\hat{c}_{\veck,\uparrow}^{\dagger}\;,\\
\hat{d}_{\veck,2}^{\dagger}&=&\hat{c}_{-\veck,\downarrow}^{\phantom{\dagger}}\;,\\
\hat{d}_{\veck,3}^{\dagger}&=&\hat{c}_{\veck+\vecQ,\uparrow}^{\dagger}\;,\\
\hat{d}_{\veck,4}^{\dagger}&=&\hat{c}_{-\veck-\vecQ,\downarrow}^{\phantom{\dagger}}\;,
\end{eqnarray}
for all $\veck\in {\mathcal B}_0$. With these operators, we
 may write the operators in~(\ref{yxc})-(\ref{yxc2}) as
 \begin{eqnarray}\label{yxc3}
\hat{\Delta}_0&=&\frac{1}{L}\sum_{\veck\in{\mathcal B}_0}\left ( \hat{d}_{\veck,1}^{\dagger}
 \hat{d}_{\veck,2}
   +
 \hat{d}_{\veck,3}^{\dagger}
 \hat{d}_{\veck,4} 
\right)\;,
\\
\opdelpi&=&\frac{1}{L}\sum_{\veck\in{\mathcal B}_0}\left ( \hat{d}_{\veck,1}^{\dagger}
 \hat{d}_{\veck,4} + \hat{d}_{\veck,3}^{\dagger}
 \hat{d}_{\veck,2}
\right)\;,
 \\\label{yxc4}
   \opnpi  &=&\frac{1}{2 L}\sum_{\veck\in{\mathcal B}_0}\sum_{j=0}^1
(-1)^{j}\\\nonumber
&&\;\;\;\;\times \left (
  \hat{d}_{\veck,1+j}^{\dagger}
 \hat{d}_{\veck,3+j}  + \hat{d}_{\veck,3+j}^{\dagger}
 \hat{d}_{\veck,1+j}
  \hat{\Delta}_0\right)\;.
\end{eqnarray}
For the 
 single particle energies in~(\ref{qwe}) we obtain
\begin{eqnarray}\nonumber
\sum_{\veck,\sigma}
\varepsilon_{\veck}
\langle 
\hat{c}_{\veck,\sigma}^{\dagger}
\hat{c}_{\veck,\sigma}^{\phantom{\dagger}} \rangle&=&
\sum_{\veck\in{\mathcal B}_0} \varepsilon_{\veck} \Big(
\langle \hat{d}_{\veck,1}^{\dagger}
 \hat{d}_{\veck,1}
  \rangle-\langle \hat{d}_{\veck,2}^{\dagger}
 \hat{d}_{\veck,2}
  \rangle
\\\label{yxc5}
&&-\langle \hat{d}_{\veck,3}^{\dagger}
 \hat{d}_{\veck,3}
  \rangle+\langle \hat{d}_{\veck,4}^{\dagger}
 \hat{d}_{\veck,4}
  \rangle
\Big)\;,\\\nonumber
\mu \sum_{\veck,\sigma}
\langle 
\hat{c}_{\veck,\sigma}^{\dagger}
\hat{c}_{\veck,\sigma}^{\phantom{\dagger}} \rangle&=&
\mu\sum_{\veck\in{\mathcal B}_0}  \Big(
\langle \hat{d}_{\veck,1}^{\dagger}
 \hat{d}_{\veck,1}
  \rangle-\langle \hat{d}_{\veck,2}^{\dagger}
 \hat{d}_{\veck,2}
  \rangle
\\\label{yxc6b}
&&\langle \hat{d}_{\veck,3}^{\dagger}
 \hat{d}_{\veck,3}
  \rangle-\langle \hat{d}_{\veck,4}^{\dagger}
 \hat{d}_{\veck,4}
  \rangle
\Big)\;,
\end{eqnarray}
With equations~(\ref{qwe}), (\ref{yxc3})-(\ref{yxc6b}), we have 
 determined the energy
\begin{equation}
\langle \hat{K}  \rangle
\equiv E(\tilde{\rho})
\end{equation}
 as a function of the single-particle 
density matrix $\tilde{\rho}$.
This matrix is diagonal with respect to
 $\veck$ and therefore determined by the four-dimensional matrices
\begin{equation}
\rho_{\veck;\gamma,\gamma'}=
\langle \hat{d}_{\veck,\gamma}^{\dagger} \hat{d}_{\veck,\gamma'} \rangle\;,
\end{equation}
for each $\veck\in {\mathcal B}_0$.

\subsection{Out of equilibrium dynamics}\label{theory2}
The time dependence of the single-particle density matrix 
 $\tilde{\rho}$ is governed by the well-known 
 differential equation~\cite{buenemann2013a,ring1980,blaizot1986}
\begin{equation}\label{faw}
\rmi \dot{\tilde{\rho}}=[\tilde{h},\tilde{\rho}]\;,
\end{equation}
where $\tilde{h}$ is also diagonal with respect to $\veck$ and defined as
\begin{equation}
\tilde{h}_{\veck;\gamma,\gamma'}\equiv 
\frac{\partial}{\partial \rho_{\veck;\gamma',\gamma}}
E(\tilde{\rho})\;.
\end{equation}
  Explicitly, $\tilde{h}_{\veck;\gamma,\gamma'}$ is given by the four-dimensional
 matrix
\begin{equation}\label{aer}
\tilde{h}_{\veck}=
 \left(\begin{array}{cccc}
 \varepsilon_{\veck}-\mu & -\eta_{\rm sc} & -\eta_{\rm co}&-\delta \eta_{\rm sc}\\
 -\eta_{\rm sc}^* & -\varepsilon_{\veck} +\mu & -\delta\eta_{\rm sc}^* &\eta_{\rm co}  \\
-\eta_{\rm co}  &-\delta \eta_{\rm sc} &  -\varepsilon_{\veck}-\mu &- \eta_{\rm sc}\\
-\delta \eta_{\rm sc}^* &\eta_{\rm co} & - \eta_{\rm sc}^*  &  \varepsilon_{\veck}+\mu
\end{array} \right)\;,
\end{equation}
with the four eigenvalues
\begin{eqnarray}\label{eq:evals}
  E_\pm^2(\veck) &=& \delta \eta_{\rm sc}^2+\varepsilon_{\veck}^2+\mu^2+\eta_{\rm sc}^2
  + \eta_{\rm co}^2 \\ &\pm& 2\sqrt{\varepsilon_{\veck}^2(\delta \eta_{\rm sc}^2+\mu^2)
    +(\delta \eta_{\rm sc}\eta_{\rm sc}+\mu\eta_{\rm co})^2}\nonumber\,.
  \end{eqnarray}
Here, the `fields'  
\begin{eqnarray}\label{yvb1}
\eta_{\rm sc}&=&U\Delta_0\;, \\
\eta_{\rm co}&=&U\npi-\alpha_{\vecQ}/2\;, \\ \label{yvb2}
\delta \eta_{\rm sc}&=&U \delpi\;,
\end{eqnarray}
 are, through (\ref{yxc})-(\ref{yxc4}),
 time-dependent functions that need to be determined self-consistently.
The calculation of these fields is simplified significantly by 
 our  $A$/$B$ lattice structure and the resulting 
property~(\ref{disp}) of the dispersion
 relation. It allows us to replace all momentum-space integrals 
 by energy integrals, e.g., 
\begin{equation}
\Delta_0=\int {\rm d}\varepsilon D(\varepsilon)
\left (\langle \hat{d}_{\varepsilon,1}^{\dagger}
 \hat{d}_{\varepsilon,2}
  \rangle +
\langle \hat{d}_{\varepsilon,3}^{\dagger}
 \hat{d}_{\varepsilon,4}
  \rangle 
\right)\;,
\end{equation}
where we introduced the (bare) density of states
\begin{equation}
D(\varepsilon)=\frac{1}{L}\sum_{\veck \in{\mathcal B}_0}\delta(\varepsilon-\varepsilon_k)\;.
\end{equation}
In the following, we will work with the semi-elliptic density of states
\begin{equation}\label{dep}
D(\varepsilon)=\frac{2}{\pi J^{2}}\sqrt{J^2-\varepsilon^2}\;\;\;\;(\varepsilon\leq 0)\;\;\;,
\end{equation}
in which $J$ sets the energy scale of our model. We note in passing that 
 in test calculations we observed only  minor quantitative changes
   of the results when we replace~(\ref{dep}) by the more realistic 
density of states of a 
 two-dimensional square lattice with nearest-neighbor hopping.

Since $\tilde{\rho}$ and $\tilde{h}$ are block diagonal with respect to $\veck$ 
 (or from now on $\varepsilon$), we need to solve the
four differential equations
\begin{equation}\label{jkl22}
\rmi \dot{\tilde{\rho}}_{\varepsilon}
=[\tilde{h}_{\varepsilon},\tilde{\rho}_{\varepsilon}]
\end{equation}
for each $\varepsilon$. It is clear that, in our numerical solution,
 we have to discretize the energy interval $-J\leq \varepsilon \leq 0$ 
and solve~(\ref{jkl22}) 
for a finite
 number $N_{\rm disk}$ of energy points $\varepsilon_i$. We found a number 
 of $N_{\rm disk}=10^4$ to be sufficiently accurate. 
 
The differential equations~(\ref{jkl22}) cannot be solved analytically 
because the fields in~(\ref{aer}) are unknown time-dependent functions.
Hence we use the numerical Adams--Bashforth~\cite{bashforth1883}
 method to $4$-th order.
After each time-step, we have to recalculate the 
fields (\ref{yxc})-(\ref{yxc2}).

It is worth mentioning that the total particle number, although no
conserved quantity in the BCS approximation due to the breaking
of $U(1)$ symmetry, is 
conserved in the time evolution described by Eq.~(\ref{faw}).
   Only numerical errors could lead to 
 an error in the particle number as a function of time. 
 This error, however, was found to be negligible for the time periods 
which we are interested in.

\subsection{Simulation of pump-and-probe experiments}\label{qyx}
In a typical pump-and-probe experiment, the system under investigation
 is in its ground state $|\phi_0\rangle$ 
(or in equilibrium at finite temperatures) 
 at some time $t=-T$.  
 Then, in the time interval $t\in (-T,0)$,  a `large' pump field is applied 
 that drives the system into some non-equilibrium state $|\phi(0) \rangle$
 at time $t=0$. The further time evolution $|\phi(t) \rangle$ 
 of this state follows from a solution of the time-dependent 
Schr\"odinger equation for the Hamiltonian $\hat{H}$ of the system (in our case 
 $\hat{K}$). 

One is interested in the response of the system to a 
 `small' probe pulse of the form 
\begin{equation}\label{aqw}
\hat{V}= \sin{(\omega t)}\Omega(t) \hat{A} \equiv f(t) \hat{A} 
\end{equation} 
that is applied at times $t>0$. Here,  $\Omega(t)$ is an envelope 
 function and $\hat{A}$ some operator, e.g., 
\begin{equation}\label{aqw2}
 \hat{A}=\hat{\Delta}_0 \;\; {\rm or}\;\; 
\hat{A}=\opnpi\;,
  \end{equation}
i.e., the operators that describe the the SC amplitude or the charge modulation. 
 The 
 wave function 
$|\phi_{\rm p}(t) \rangle$, in the presence of the probe pulse, differs from 
 $|\phi(t) \rangle$ and so does the expectation value of $\hat{A}$.
 Hence, we may define
 \begin{equation}\label{qop}
\delta A(t)\equiv  \langle\hat{A} \rangle_{\phi_{\rm p}(t)}- 
  \langle\hat{A} \rangle_{\phi(t)}
\end{equation} 
 as a measure for the impact of $\hat{V}$ on the observable $\hat{A}$. 

Without the pump pulse, $\delta A(t)$ is usually calculated by means of
 the Kubo formula~\cite{kubo1957,kubo1959}
\begin{equation}\label{jkl}
\delta A(t)=\int_{0}^t{\rm d} t' \chi_{A,A}(t,t')f(t')
\end{equation} 
 with the (retarded) Green's function
\begin{equation}\label{jkl2}
 \chi_{A,A}(t,t')\equiv-\rmi \theta(t-t') \langle\phi(0)|
[\hat{A}_{\rm I}(t),\hat{A}_{\rm I}(t')]
|\phi(0)
\rangle
\end{equation} 
where $\hat{A}_{\rm I}(t)=e^{\rmi \hat{H}t}\hat{A} e^{-\rmi \hat{H}t}$ 
is the interaction representation of  $\hat{A}$ (i.e., $\hat{H}$ is defined
{\it without} the probe pulse). Note that, in 
a linear-response
 approximation,  Eqs.~(\ref{jkl}),(\ref{jkl2}) are equally valid if 
 $| \phi(0) \rangle$ is the excited state induced by the pump pulse.
 However, the full two-time response function
 $\chi_{A,A}(t,t')$ is needed here, instead of $\chi_{A,A}(t-t')$ when the
 perturbation is applied to the ground state. In order to show this,
we introduce the eigenstates $ |n\rangle$ and energies $E_n$
 of  $\hat{H}$
 and   the expansion
\begin{equation}
| \phi(0) \rangle=\sum_m \varphi_m | m\rangle 
\end{equation} 
of the initial state. With these, we can write, e.g., the first part
 of the commutator in~(\ref{jkl2}) as
\begin{eqnarray}
&&\langle\phi(0)|
\hat{A}_{\rm H}(t)\cdot\hat{A}_{\rm H}(t')
|\phi(0)
\rangle\\\nonumber
&&=
\sum_{m,m',n}
\varphi^*_m\varphi_{m'}
A_{m,n}A_{n,m'}
e^{-{\rm i}E_n(t-t')}
e^{{\rm i}E_m t}
e^{-{\rm i}E_{m'} t'}\;,\\
&&A_{m,n}\equiv \langle m|\hat{A}|n\rangle\;.
\end{eqnarray} 
Obviously, this quantity is a function of $t-t'$ only when $m=m'$, i.e., when 
$| \phi(0) \rangle$ is an eigenstate of $\hat{H}$.
 The (non equilibrium) Greens function therefore has
  unusual properties as becomes clear from the equivalent of a
 Lehmann representation: 
  First,  we can perform a Fourier transform of
(\ref{jkl2}) with respect to $\tau\equiv t-t'$ while explicitly
keeping the response time $t$
\begin{eqnarray}  \label{10.1300} 
  \chi_{A,A}(\omega,t)&=& \sum_{n,m,m'} \varphi^*_m\varphi_{m'}e^{\rmi(E_m-E_{m'})t}
  \\\nonumber
&&\times\left[ \frac{A_{mn}A_{nm'}}{\omega-E_{n,m'}+\rmi\delta }-\frac{A_{mn}A_{nm'}}{%
\omega+E_{n,m}+\rmi\delta }\right]\;,   
\end{eqnarray} 
where $E_{n,m}\equiv E_{n}- E_{m}$.
Upon further defining a `long-time' response average \cite{rossini14}
\begin{equation}
\widetilde{\chi}_{A,A}(\omega)=\frac{1}{T'} \int_0^{T'} dt \chi_{A,A}(\omega,t)
\end{equation}
one obtains for $T'\to \infty$
\begin{eqnarray}  \label{10.1301} 
  \widetilde{\chi}_{A,A}(\omega)&\to& \sum_{n,m} |\varphi_m|^2
  \\\nonumber
&&\times\left[ \frac{A_{mn}A_{nm}}{\omega-E_{n,m}+\rmi\delta }-\frac{A_{mn}A_{nm}}{%
\omega+E_{n,m}+\rmi\delta }\right]\;.   
\end{eqnarray} 
This has a similar structure as the equilibrium response function but for the
factor $|\varphi_m|^2$ which describes the admixture of excited states induced
by the pump pulse. It has been shown \cite{rossini14} that, for
a number of cases, Eq.~(\ref{10.1301}) is similar to the
equilibrium response when the $|\varphi_m|^2$ are replaced by
Boltzmann weights. 
In this spirit we will later compare the numerically obtained
response, using Eqs. (\ref{jkl22},\ref{aqw}), with an equilibrium
response function for a non-zero effective temperature.

In equilibrium, the linear-response assumption of 
Eqs.~(\ref{jkl}),(\ref{jkl2}) is justified
 because a small perturbation will normally lead to a small 
response of a 
system that is in its stationary ground state and sufficiently far away
from an instability. The situation is obviously 
different when the system is in a non-equilibrium state due to 
 a pump pulse and it is not clear to what extend 
(\ref{jkl}),(\ref{jkl2}) are still applicable. Since the TDHF method 
 that we  use in this work does not rely on the  
 linear-response assumption we are able to assess its validity in  
  pump-and-probe situations.

Relevant pump pulses are of the form given in 
 (\ref{aqw})-(\ref{aqw2}). Hence, to define them, we have to specify the pulse
 frequency as well as the shape and duration $T$ of the pump pulse
$ \Omega(t)$. 
 These tunable quantities would come on top of the system parameters 
$U$, $n$, $\alpha_{\vecQ}$ and the probe frequency  $\omega$. To limit  
 the total number of such parameters, 
and since we are not addressing any specific 
experiment, we prefer to set up the initial out-equilibrium states not 
 through a pump pulse but by 
 varying the three initial 
 fields $\{\eta^0_{\nu}\}\equiv
\{\eta^0_{\rm sc}, \delta \eta^0_{\rm sc}, \eta^0_{\rm co}\}$  away 
from their ground-state 
values $\eta^{\rm gs}_{\nu}$. 

For a given set of  initial 
fields $\eta^0_{\nu}$ (at time $t=0$) and a given probe pulse
Eq.~(\ref{aqw}) we solve Eq.~(\ref{faw})
 numerically over a certain time period $\Delta t$, typically 
 $\Delta t=1000/J$. With this
 solution, we determine both expectation values on the r.h.s.~of~(\ref{qop}) 
and hereby the fluctuations $\delta A(t)$.
Note that, for a non-Hermition operator $\hat{A}$ (e.g., $\Delta_0$) 
the latter contain both amplitude and phase contributions, i.e.,
\begin{eqnarray}
  |\delta A(t)|^2&=&|\langle\hat{A} \rangle_{\phi_{\rm p}(t)}|^2+ 
  |\langle\hat{A} \rangle_{\phi(t)}|^2  \\
  &-& 2|\langle\hat{A} \rangle_{\phi_{\rm p}(t)}| 
  |\langle\hat{A} \rangle_{\phi(t)}|\cos(\Phi_{\rm p}(t)-\Phi_0(t)) \nonumber
\end{eqnarray}
where $\Phi_{{\rm p}(0)}(t)$ denotes the phase of 
$\langle\hat{A} \rangle_{\phi_{{\rm p}(0)}(t)}$ with (without)
the probe pulse.
 As a measure 
 for the impact of the probe pulse, we define
\begin{equation}\label{rtz1}
\langle \delta A  \rangle=\frac{1}{\Delta t}\int_0^{\Delta t} {\rm d}t |\delta A(t)|\,.
\end{equation}
This quantity will be considered as a function of $\omega$, the pulse frequency
 in~(\ref{aqw}), where, for simplicity, we set $\Omega(t)=\Omega_o=10^{-5}$.
 In cases where the experiment measures only the amplitude with and without the
 probe pulse, in particular in connection with SC,  it is also useful
 to define the response quantity 
 \begin{equation}\label{rtz2}
\langle \delta| \Delta_0 | \rangle\equiv
\frac{1}{\Delta t} \int_0^{\Delta t} {\rm d}t \Big| | \langle \hat{\Delta}_0  \rangle_{\Phi_{\rm p}(t)} 
|- | \langle \hat{\Delta}_0  \rangle_{\Phi(t)} |     \Big |.
\end{equation}
 If the response occurs on top of a non-equilibrium state with 
 $ | \langle \hat{\Delta}_0  \rangle_{\Phi(t)} | \gg |\delta\Delta_0(t)|$
 one can expand Eq.~(\ref{rtz1})
 \begin{equation}\label{eq:d0c}
   \langle \delta| \Delta_0 | \rangle =
\frac{1}{2 \Delta t}   \int_0^{\Delta t} {\rm d}t \Big| \delta\Delta_0(t){\rm e}^{-i\phi(t)}
   +\delta\Delta^*_0(t){\rm e}^{i\phi(t)}\Big|
 \end{equation}
 where $\phi(t)$ denotes the phase of $\langle \hat{\Delta}_0  \rangle_{\Phi(t)}$.
 For a linear phase change $\phi(t)=\Omega_P t$ and upon performing the
 same response average which lead to Eq.~(\ref{10.1301}) one therefore
 expects an amplitude response $\langle \delta| \Delta_0 | \rangle$
 with a two-peak structure corresponding to the absorption and emission
 of a phase mode.
 
\section{Ground-state properties}\label{gs} 
We start with a consideration of our model's ground-state properties.
The technical details of the numerical minimization are briefly 
outlined in Appendix~\ref{app1}.

\begin{figure}[bt]
\includegraphics[width=8cm]{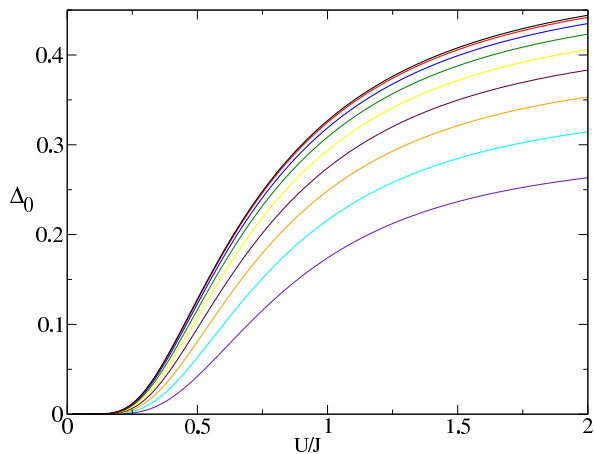} 
\caption{Superconducting order parameter $\Delta_0$ 
in the  superconducting ground state as a function of $U/J$ 
for band fillings $n=1,0.9,0.8,0.7,0.6,0.5,0.4,0.3,0.2$ (in descending order).}
\label{Fig:fig1}
\end{figure}

We first consider the case without external CDW field
$\alpha_{\vecQ}=0$. As well known,\cite{micnas90} 
a purely superconducting phase, i.e., with 
$\eta_{\rm co}= \delta \eta_{\rm sc}=0$ is stable for all $U>0$. 
We display the resulting  (real) order parameter $\Delta_0$ 
in the  superconducting ground state
in Fig.~\ref{Fig:fig1} as a function of $U/J$ 
for various band fillings. 
 Note that, due to particle-hole symmetry, it is sufficient 
 throughout this work
  to consider only results below half filling, $n\equiv 2n_0\leq 1$. 

\begin{figure}[b] 
\includegraphics[width=8cm]{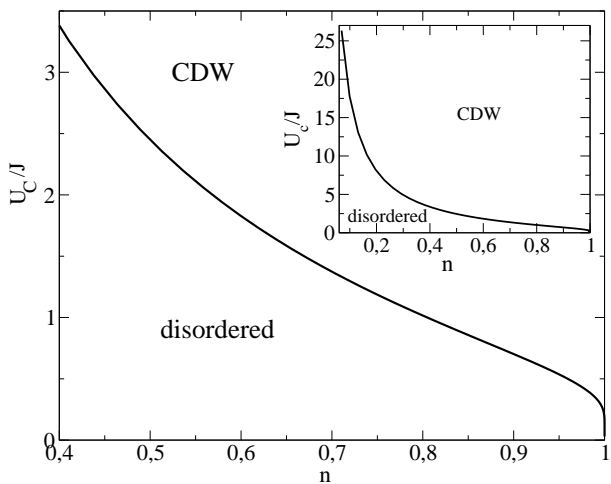} 
\caption{Phase diagram for a pure charge-ordered phase (i.e., without SC)
  as a function of $n$;
  Inset: the same result for a larger range of densities $n$.}
\label{phase_diag_et.eps}
\end{figure}

\begin{figure}[bt] 
\includegraphics[width=8cm]{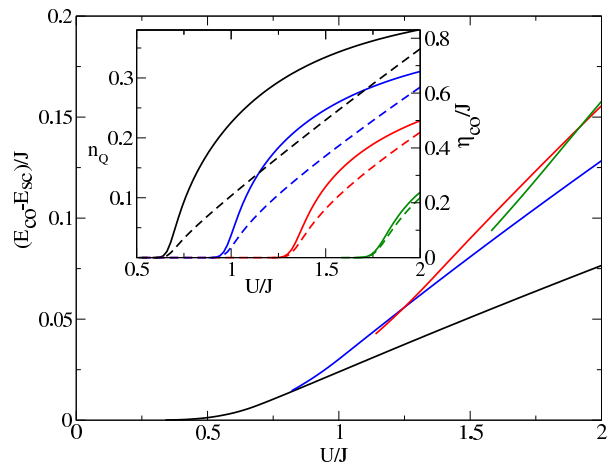} 
\caption{Energy difference between the  superconducting and 
the charge-ordered phase for $n=0.9$ (black), $0.8$ (blue), $0.7$ (red)
 $0.6$ (green) as a function of $U/J$; Inset: Corresponding results
 for the  charge order parameter $\npi$ (solid lines) and the 
 fields $\eta_{\rm co}$ (dashed lines)}.
\label{en_diff.eps}
\end{figure}

   Formally, a purely charge-ordered state can be induced
   from a second order instability 
above some critical value $U_{\rm C}$ that depends on the particle density 
$n$.
 An analytical analysis of the energy functional reveals that 
 $U_{\rm C}$
 is given by
\begin{equation}
 U_{\rm C}=\left[\int^{J}_{|\mu|} {\rm d}\varepsilon \frac{D(\varepsilon)}{\varepsilon}\right]^{-1},
\end{equation}
which is displayed as a function of $n$ in Fig.~\ref{phase_diag_et.eps}. 
    At half filling, due to perfect nesting, $U_{\rm C}$ goes to zero and
    $n=1$ corresponds also to the peculiar situation where CDW and
    SC ground states are energetically degenerate.
 For all densities $n$ away from half filling, the SC phase
 is lower in energy than the charge-ordered phase. This can be seen in 
 Fig.~\ref{en_diff.eps} where we show the
energy difference between the superconducting and 
the charge-ordered phase as a function of $U/J$ for various values 
 of $n$. The inset of this figure displays the corresponding 
 charge order parameters $\npi$ and fields $\eta_{\rm co}$.
 It should be mentioned that here we are restricted to a commensurate
 [i.e., ${\bf Q}=(\pi,\pi)]$ CDW whereas away from half-filling incommensurate
 charge orders would be energetically more stable albeit still above the SC
 ground-state energy.

For Coulomb parameters of $U$ where both phases are stable, it is conceivable
  that a coexistence phase, i.e.,  with $\Delta_0$ and $\npi$ both 
non-zero,
 has an even lower  energy. In our numerical analysis, however, 
we found such a phase to be always  
 higher in energy than a pure superconducting one. 

\begin{figure}[bt] 
\includegraphics[width=8cm]{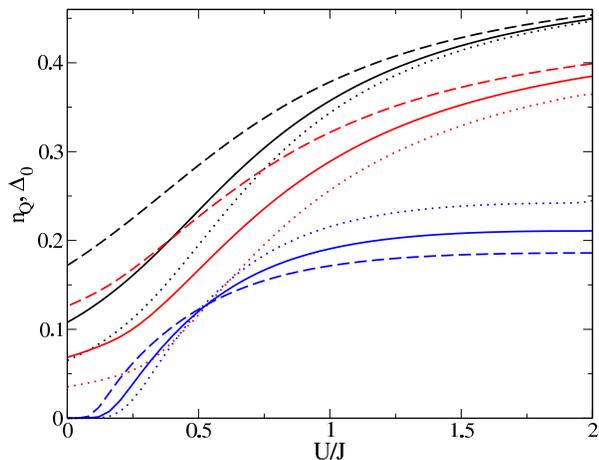} 
\caption{Order parameters $\npi$ (black ($n=1.0$) and red ($n=0.9$))  
and $\Delta_0$ (blue ($n=0.9$))  as a function 
 of $U/J$ for $\alpha_{\vecQ}=0.05,0.1,0.2$ (dotted, solid,dashed). 
 Note that $\Delta_0=0$ for $n=1.0$.}
\label{dn-del-of_U.eps}
\end{figure}

\begin{figure}[bt] 
\includegraphics[width=8cm]{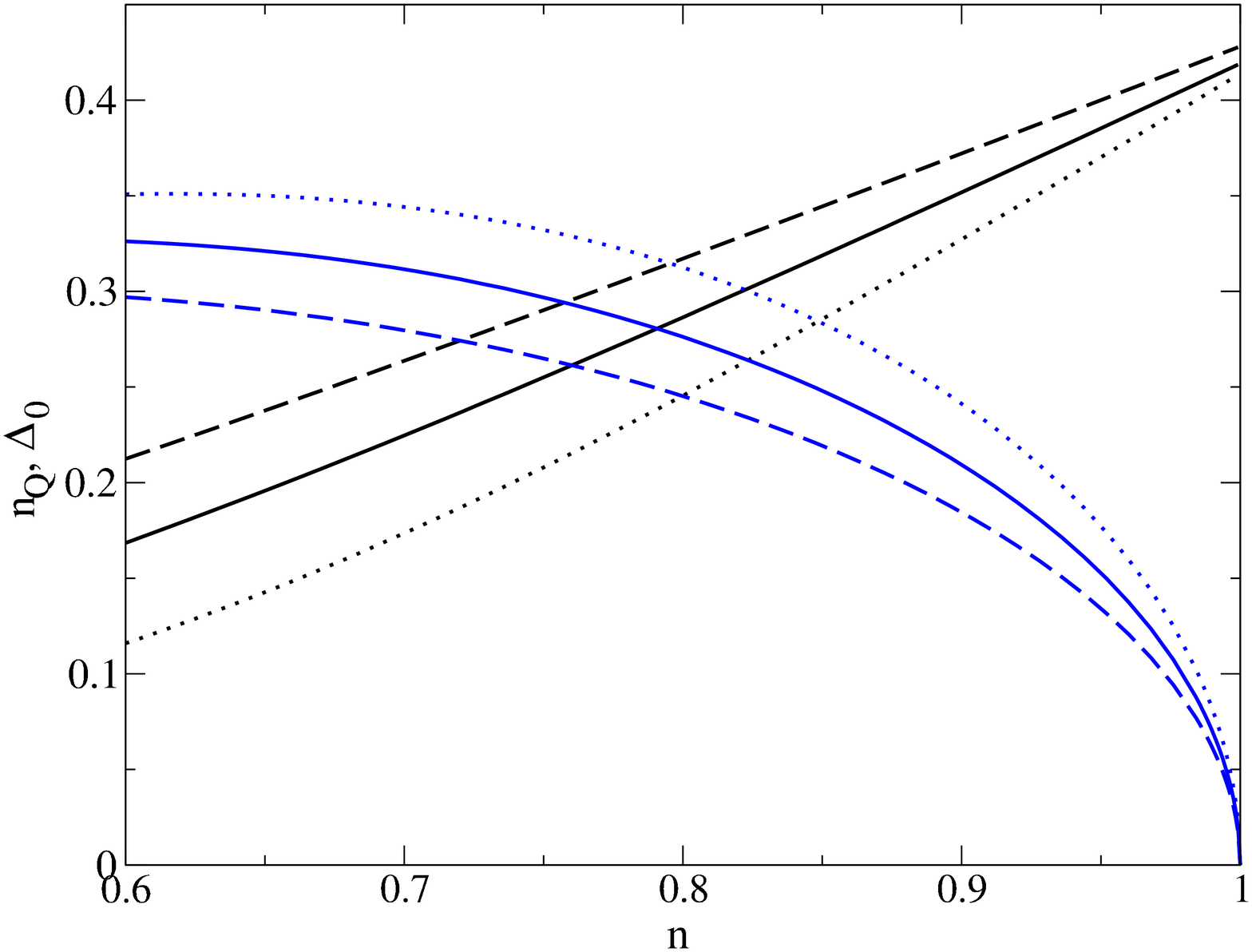} 
\caption{Order parameters $\npi$ (black)  
and $\Delta_0$ (blue)  as a function 
 of charge concentration $n$ for $U/J=1.5$ and 
$\alpha_{\vecQ}=0.05,0.1,0.2$ (dotted, solid,dashed)}
\label{dn-del-of_n.eps}
\end{figure}

 In order to stabilize charge order and to allow for situations
 with both order parameters finite, 
 we introduce a non-zero charge-order field
 $\alpha_{\vecQ}$, see Eq.~\ref{rta}. In its presence, a purely charge-ordered phase is obviously  
 most favorable at half filling. Away from half filling, 
both order parameters $\npi$ and  $\Delta_0$ are
 non-zero in the ground state. We show the typical behavior of both 
 parameters as a function of $U$ for $n=1.0$ and $n=0.9$ in 
Fig.~\ref{dn-del-of_U.eps}.
The doping dependence of both quantities is displayed in 
Fig.~\ref{dn-del-of_n.eps}. Note that in approaching half filling, the 
 pairing order parameter is non-analytic, $\Delta_0\sim \sqrt{1-n}$. 

In the region $U/J \lesssim 0.5$, the superconducting order parameter shows
 a somewhat unexpected behavior because it gets larger when 
 $\alpha_{\vecQ}$ is increased. In this regime the chemical
 potential falls in the range where the CDW opening induces a $1/\sqrt{\omega}$ enhancement of the DOS. For larger $U/J$ the SC order parameter gets suppressed
 by the CDW scattering as expected.
 
\section{Out-of-equilibrium dynamics}\label{ooed}
 In this chapter, we discuss the out-of-equilibrium dynamics of our model.
 In the first section, we consider situations without an external 
time-dependent  perturbation (`quantum-quench problems'), where the time 
 dependence results from an initial state that is not the ground state. 
 The problem of `pump and probe experiments' is studied in the second 
 section. 
\subsection{Quantum quench problems}\label{qqp}
We first consider the dynamics of our system 
 that evolves from different initial density matrices. The latter are 
 determined by the initial parameters  
$\eta^0_{\rm sc}, \eta^0_{\rm co}, \delta \eta^0_{\rm sc}$. 
In the following we shall always change these 
 parameters relative to their ground state values,
$\eta^0_{\nu}= \gamma \eta^{\rm gs}_{\nu}$. 
There are three different
 situations, (i) a purely  
charge-ordered phase  ($\eta^0_{\rm sc}= \delta \eta^0_{\rm sc}=0$),  
 (ii) a superconducting phase 
($\eta^0_{\rm co}= \delta \eta^0_{\rm sc}=0$), and (iii) 
 a  coexistence phase of both orders where all $\eta^{\rm gs}_{\nu}$ are
 non-zero. Note that our dynamics does {\it not} evolve from
 a Hamiltonian with `quenched' interaction parameters but only
 from a `quenched' density matrix. This is different from previous studies
 \cite{manmana07,schiro2010,schiro2011,eckstein10,hamerla13,hamerla14,bauer15}
 where in the context of the (repulsive) Hubbard model the local
 interaction $U$ is set to a different value at $t=0$.

\begin{figure}[hhh!]
\includegraphics[width=8cm]{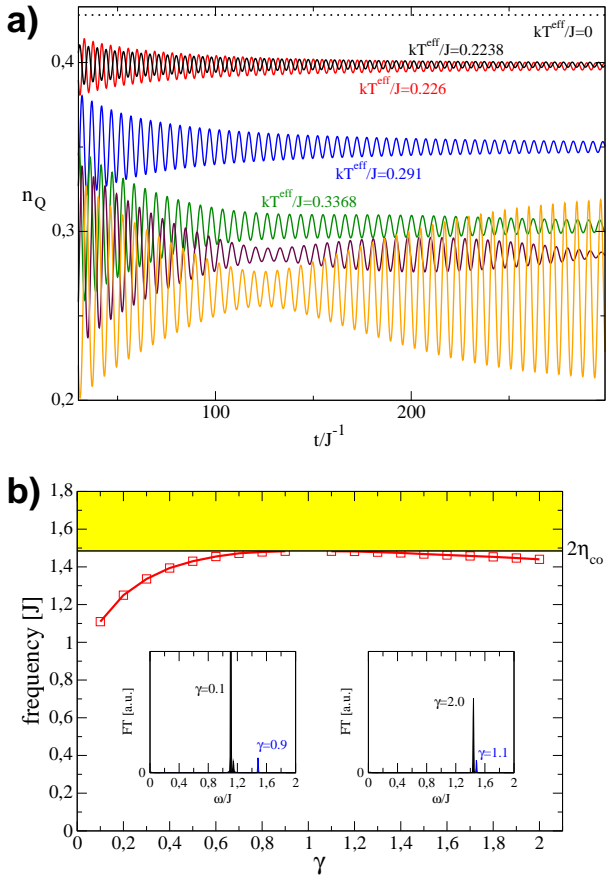}
\caption
{a) Charge-order parameter $n_{\vecQ}$ as a function of time $t$ 
  for $U/J=1.5$, $n=1.0$, 
$\alpha_{\vecQ}=0.2$,
and scaling factors $\gamma=3.0$ (black, solid), $\gamma=1.0$ (black, dotted), 
$\gamma=0.4$ (red), $\gamma=0.2$ (blue), $\gamma=0.1$ (green), 
$\gamma=0.075$ (maroon),$\gamma=0.05$ (orange). For moderate $\gamma$ the
envelope function can be fitted by $n_{\vecQ}^{\infty}\pm C/\sqrt{t}$ (dashed)
and the effective temperature $T^{\rm eff}$, as indicated in the lower panel, is defined as the temperature
for which an equilibrium calculation yields $n_{\vecQ}^{\infty}$ for otherwise
the same parameters. b) Frequency of the Fourier peaks as a function of
$\gamma$. The insets detail the Fourier spectra for selected $\gamma$ values.
For $\gamma\to 1$ the excitations approach the CDW gap $2\eta_{CO}$.}
\label{et-u=1.5-n=1.0-etabare=0.2.eps}
\end{figure}

We start with a consideration of case (i). In 
Fig.~\ref{et-u=1.5-n=1.0-etabare=0.2.eps}a we show 
$n_{\vecQ}$ as a function of time $t$ for $U/J=1.5$, $n=1.0$, $\alpha_{\vecQ}=0.2$ 
 and several scaling factors $\gamma$ between $0.05$ and $3$. 
 The time-dependence of $n_{\vecQ}$ has the generic structure
 expected for mean-field order parameters \cite{altshuler06}
which  previously has been
 discussed in the context of SC
 \cite{volkov73,barankov04,papenkort07,papenkort08,krull14} or
 antiferromagnetism. \cite{tsuji13} This includes a
 $\cos(2\eta_{CO}^\infty t)/\sqrt{\eta_{CO}^\infty t}$ relaxation
 of the amplitude towards a stationary value $n_Q^\infty$ which appears as a
 consequence
 of a `dephasing' between the individual contributions of the
 scattering processes ${\bf k} \to {\bf k}+{\bf Q}$ to the order parameter.
Close to $\gamma=1$ the oscillatory frequency
 is determined by the amplitude excitations across the CDW gap $2\eta_{CO}$,
 which soften with increasing deviation from the equilibrium state
 towards a values $\eta_{CO}^\infty=U n_{\vecQ}^\infty+\alpha_{\vecQ}/2$ for
 $t\to \infty$. 
 The difference between this (reduced) stationary value as compared
 to the equilibrium result at $\gamma=1$ may be interpreted in terms
 of a population of excited HF states via an effective finite temperature
 $T^{\rm eff}$. The latter is defined by the condition that the  Fermi-Dirac 
distribution
 for $T^{\rm eff}$ leads to an equilibrium expectation value of $\hat{n}_{\vecQ}$
 that equals $n_{\vecQ}^{\infty}$. The values of these effective temperatures
  are given close to the
  corresponding curves in Fig.~\ref{et-u=1.5-n=1.0-etabare=0.2.eps}a.
  
 The behavior of the dynamics changes
 for  $\gamma< 0.1$ where oscillations on much longer time scales emerge. 
 In this regime one should go to very large times in order to
   obtain sensible results which conflicts with the stability of integration.
   We therefore abstain from an investigation of the extreme non-equilibrium
 regime.

 The dependence of the CDW amplitude excitations on $\gamma$, obtained from
 the Fourier spectra of $n_{\vecQ}(t)$, is summarized in 
 Fig.~\ref{et-u=1.5-n=1.0-etabare=0.2.eps}b. Close to $\gamma=1$ the
 linear-response dynamics of the CDW amplitude is described by a peak at
 $\Omega=2\eta_{CO}$ with small intensity (cf. spectra for $\gamma=0.9$,
$1.1$  in the
 insets) due to the strong mixing
 with quasiparticle excitations (indicated by the yellow shaded area).
 Upon increasing the
 non-equilibrium situation (i.e., $|\gamma-1|$) the excitations soften
 and move inside the (equilibrium) CDW gap. 
  Note that the equilibrium value of $n_{\vecQ}\approx 0.43$ in Fig.~\ref{et-u=1.5-n=1.0-etabare=0.2.eps} is close to the maximum value $n_{\vecQ}= 0.5$ for
 a CDW at half-filling. Therefore one needs a large $\gamma>1$ in order
 to approximately obtain the same stationary value as for $\gamma=0.4$
 and consequently also the CDW excitations are not symmetric with
 respect to $\gamma=1$.
 

\begin{figure}[ttt!]
%
%
\includegraphics[width=7.5cm]{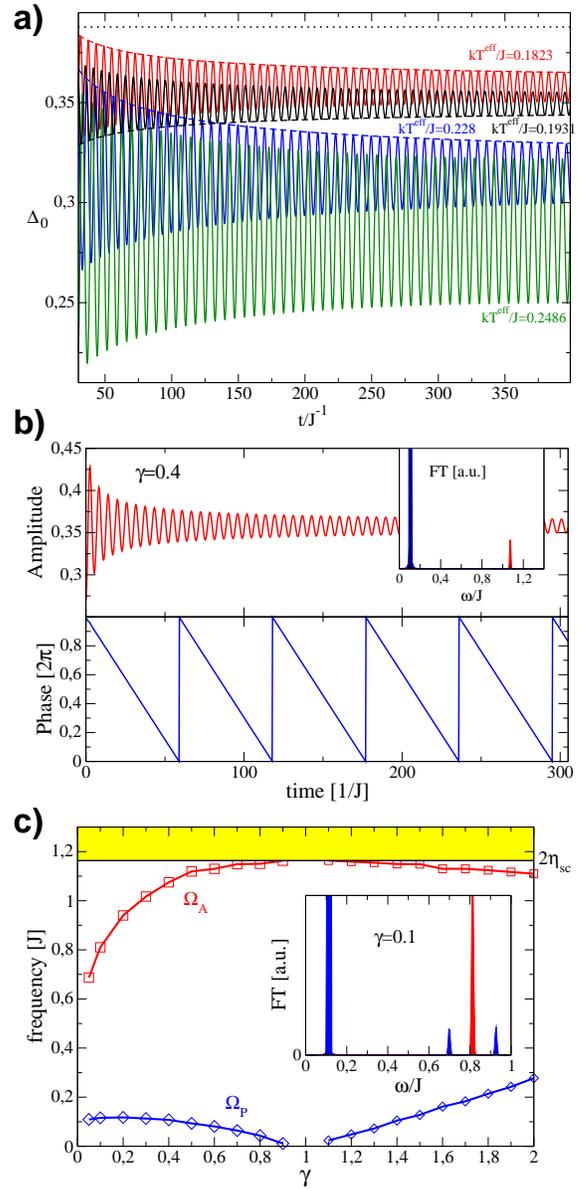}
\caption
{a) Absolute value of the pairing order parameter $|\Delta_0|$ as a function if time $t$  for $U/J=1.5$, $n=0.7$,
and scaling factors $\gamma=3.0$ (black, solid), $\gamma=1.0$ (black, dotted), 
$\gamma=0.4$ (red), $\gamma=0.2$ (blue), $\gamma=0.125$ (green). Staggered
CDW field $\alpha_{\bf Q}=0$. For moderate $\gamma$ the
envelope function can be fitted by $n_{\vecQ}^{\infty}\pm C/\sqrt{t}$ (dashed)
and the effective temperature $T^{\rm eff}$, as indicated adjacent to the curves,
is defined as the temperature
for which an equilibrium calculation yields $n_{\vecQ}^{\infty}$ for otherwise
the same parameters.
b) Time evolution of amplitude and
phase for $\gamma=0.4$. The inset shows the corresponding Fourier peaks of
the main frequency (amplitude: red; phase: blue).
c) Amplitude and phase excitation frequency as a
function of the scaling parameter $\gamma$. The inset reports the
Fourier spectrum for $\gamma=0.1$ demonstrating the coupling between
amplitude and phase modes.
}
\label{del-u=1.5-n=0.7-etabare=0.0.eps}
\end{figure}

As an example for case (ii) we show in Fig.~\ref{del-u=1.5-n=0.7-etabare=0.0.eps}a the 
absolute value $|\Delta_0|$ of the pairing order parameter as a function 
of time 
for $U/J=1.5$, $n=0.7$, and several scaling factors $\gamma$.
Similar to the previous case the time dependence is a damped
oscillatory behavior $\cos(2\eta_{SC}^\infty t)/\sqrt{\eta_{SC}^\infty t}$
which approaches $\eta_{SC}^\infty$ for $t\to \infty$. The latter
can again be described by an effective temperature $T^{\rm eff}$ as indicated
adjacent to the corresponding curves.

Fig.~\ref{del-u=1.5-n=0.7-etabare=0.0.eps}b reports
the time evolution of SC amplitude and phase for $\gamma=0.4$ together
with their Fourier transforms. The gauge invariance of the TDHF
approach implies conservation of charge which would
be violated if one changes the order parameter in a BCS calculation
without adjusting the chemical potential. Here charge conservation is
obeyed in the non-equilibrium situation via the coupling to the
phase mode which appears at a finite frequency $\Omega_P$.
The dependence of both amplitude excitation $\Omega_A$ and
phase mode $\Omega_P$ on $\gamma$ is summarized in
Fig.~\ref{del-u=1.5-n=0.7-etabare=0.0.eps}c. The amplitude excitation
for the SC order parameter has essentially the same behavior as
for the charge order parameter in case (i) and develops from $\Omega_{A}=2\eta_{SC}$ at $\gamma=1$
inside the (equilibrium) SC gap upon decreasing or increasing
$\gamma$ from the equilibrium situation.
In equilibrium the phase mode $\Omega_P=0$ reflecting its
property as a Goldstone mode for the U(1) symmetry breaking.
Upon deviating from equilibrium, $\Omega_P$ moves inside the
SC gap. The inset to Fig.~\ref{del-u=1.5-n=0.7-etabare=0.0.eps}c
demonstrates the coupling between phase and amplitude excitation
which reflects as two side peaks at $\Omega_A \pm \Omega_P$ and
which gets more pronounced upon increasing non-equilibrium.
Note that such mixing is a common feature of superconductors in non-equilibrium
and has recently also been exploited for two-band systems where a coupling
between amplitude (Higgs) and Leggett modes can be induced. \cite{krull16}

\begin{figure}[bt]
\includegraphics[width=8cm]{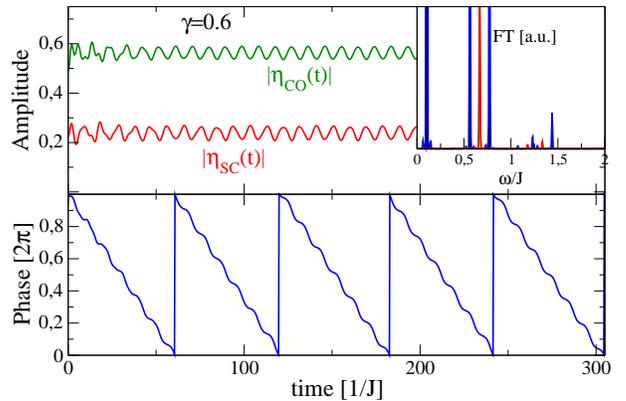}
\caption{Dynamics of CDW amplitude (green), SC amplitude (red) and
  SC phase (blue) for a system with both orders finite in the
  ground state. The inset reports the corresponding Fourier transforms (CDW and
  SC amplitude excitations occur at the same frequencies and therefore
  are undistinguishable). 
  At $t=0$ the equilibrium values of the order parameters
  are scaled with $\gamma=0.6$. $U/t=1.5$, $\alpha_{\vecQ}=0.2$, $n=0.8$.
}
\label{ampphascdw+sc}
\end{figure}

 Finally, Fig.~\ref{ampphascdw+sc} reports the amplitude and phase
 dynamics for a ground state with both SC and CDW order and scaling
 factor $\gamma=0.6$. A first obvious difference to the previous cases
 is that after a short transient response both  SC and  CDW order parameter
oscillate with constant amplitude without any signature
of relaxation. Second, the short period oscillation of CDW and SC amplitude is
now clearly imprinted onto the phase dynamics which therefore reflects
the strong coupling between amplitude and phase in this case.
As a consequence and similar to the previous case, the phase mode now also appears in the form of
 side bands in the Fourier spectrum of the SC/CDW amplitude excitation
which both appear at the same energy $\Omega_A$. Moreover, already at
$\gamma=0.6$ higher-order excitations at $2\Omega_A$ and
$2\Omega_A\pm \Omega_P$ are visible in the spectrum, though with
rather small intensity. A more detailed
inspection reveals also interference effects between phase and
amplitude oscillations at $\delta\omega=\Omega_A-6\Omega_P$
which further split the phase excitations.

\subsection{Pump and probe simulations}\label{pape}
Our general method of analyzing pump and probe situations has been described 
 in Sec.~\ref{qyx}. 
Depending on the values of $\alpha_{\vecQ}$ and $n$ there can be three different 
 equilibrium phases: a purely superconducting or charge-ordered phase, or a 
 coexistence phases of both orders. We shall consider all three cases
 separately in the following sections.
Since we can probe the response 
  of the system to both, a pairing or a charge-order field
 there are six different setups in total
 that we need to consider.

\subsubsection{Charge-order states}\label{xcv}
We first look at response functions in a charge-ordered 
 phase as it is established in the ground state, e.g.,  
for $U/J=1.5$, $n=1.0$, and  
$\alpha_{\vecQ}=0.2$. The 
 time dependence of  $n_{\vecQ}$ in this case has been shown for several values 
 of the initial scaling factor $\gamma$ in 
Fig.~\ref{et-u=1.5-n=1.0-etabare=0.2.eps}a.

We can probe the response 
  of the system to both, a pairing or a charge-order field.
Note that,  without a  probe pulse the pairing amplitude is zero.
This means  that for a pairing probe pulse,
Eq.~(\ref{rtz2}) simplifies to 
 \begin{equation}\label{rtz3}
\langle \delta| \Delta_0 | \rangle=
\frac{1}{\Delta t}
 \int_0^{\Delta t} {\rm d}t  | \langle \hat{\Delta}_0  \rangle_{\Phi_{\rm p}(t)} | .
\end{equation}

\begin{figure}[bt] 
\includegraphics[width=8cm]{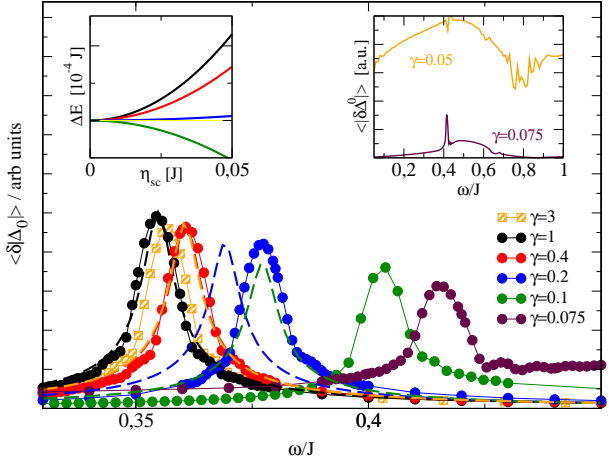} 
\caption{Pairing response  $\langle \delta| \Delta_0|\rangle $
 as a function of frequency $\omega$
for $U/J=1.5$, $n=1.0$,  $\alpha_{\vecQ}=0.2$, 
and scaling factors $\gamma=3.0$ (black, solid), $\gamma=1.0$ (black, dotted), 
$\gamma=0.4$ (red), $\gamma=0.2$ (blue), $\gamma=0.1$ (green), 
$\gamma=0.075$ (maroon). ELRT results are shown with dashed lines
and are obtained for the effective temperatures indicated
in the lower panel of Fig.~\ref{et-u=1.5-n=1.0-etabare=0.2.eps}.
Upper right inset: Same quantities 
for $\gamma=0.075$ (maroon), $\gamma=0.05$ (orange).
Upper left inset: Change $\Delta E\equiv E(\eta_{\rm sc})- E(\eta_{\rm sc}=0)$
of the 
 energy as a function of $\eta_{\rm sc}/J$ for stationary states with  
$U/J=1.5$, $n=1.0$,  $\alpha_{\vecQ}=0.2$, and expectation values 
$n_{\vecQ}\approx 0.43$ (black), $n_{\vecQ}= 0.4$ (red), $n_{\vecQ}= 0.35$ (blue), 
$n_{\vecQ}= 0.3$ (green).}
\label{del-u=1.5_n=1.0:etbare=0.2.eps}
\end{figure}

In Fig.~\ref{del-u=1.5_n=1.0:etbare=0.2.eps} we show 
$\langle \delta |\Delta_0|\rangle $, as defined in (\ref{rtz3}),
 as a function of probe frequency for $U/J=1.5$, $n=1.0$,
 $\alpha_{\vecQ}=0.2$ and
  several values of $\gamma$.
  The figure shows that the qualitativ structure of the
response does not change in the same range of $\gamma$ values,
$3\le \gamma\le 0.075$,
for which the quench dynamics [Fig.~\ref{et-u=1.5-n=1.0-etabare=0.2.eps}]
shows the `regular' damped oscillatory behavior.
Starting from the equilibrium situation $\gamma=1$ one observes a
shift of the response to higher frequencies upon increasing or
decreasing $\gamma$ without a change of the Lorentzian peak
structure. This behavior is in qualitative agreement with the expectation
from equilibrium linear-response theory (ELRT)  which is derived for the
present situation in appendix \ref{app3} and evaluated 
(dashed lines in Fig.~\ref{del-u=1.5_n=1.0:etbare=0.2.eps}) with the effective 
 temperatures $T^{\rm eff}$, introduced in Sec.~\ref{qqp}.
Within this approximation,
the pairing fluctuation, induced by the onsite-attraction $U$, generates
poles within the CDW gap $2\eta_{CO}$ (cf Fig.~\ref{poles}
in appendix \ref{app3}). 
Small values of $U/J$ (compared to $\eta_{CO}/J$) induce
in-gap states close to the upper band edge $2\eta_{CO}$ and with increasing
fluctuation strength $U/J$ the pole is shifted to lower energy
inside the CDW gap.

As expected, for $\gamma=1$ ELRT agrees with the
full TDHF result as can be seen from Fig.~\ref{del-u=1.5_n=1.0:etbare=0.2.eps}. 
A moderate non-equilibrium situation
corresponds to an initial population of excited states and in
ELRT can be modelled by a finite temperature for which the
values are given in the lower panel of Fig.~\ref{et-u=1.5-n=1.0-etabare=0.2.eps}. One finds that  upon increasing non-equilibrium,
ELRT captures the reduction of peak intensity
and the shift of the peak to higher energy, however, it under(over)estimates
the latter upon decreasing (increasing) $\gamma$ from $\gamma=1$.

Only for values of  $\gamma$ below $0.075$, 
the linear-response assumption  
 seems to break down almost instantaneously, see the inset of
 Fig.~\ref{del-u=1.5_n=1.0:etbare=0.2.eps}, where also
 the quench dynamics changes.

 In an equilibrium situation the collective excitation frequency
$\Omega$  of an observable ${\cal O}$ can be deduced from the
 curvature of the energy functional around the saddle point, i.e.,
 $\Omega^2 \sim \partial^2 E/\partial(\delta{\cal O})^2$. In this
 spirit one could argue that the hardening of
 the excitation in Fig.~(\ref{del-u=1.5_n=1.0:etbare=0.2.eps})
is due to a stiffening of the ground state energy functional  
as a function of 
the dynamical variable, in our case $\Delta_0$.
Such an interpretation, 
however, fails in our   
 out-of-equilibrium situation. This is illustrated in 
the upper left inset to Fig.~\ref{del-u=1.5_n=1.0:etbare=0.2.eps} 
where we show
 the energy change as a function of $\eta_{\rm sc}$ that is induced
 in a stationary state for values of $n_{\vecQ}$ that equal those
 in the long time for $\gamma=3.0,0.4,0.2,0.1$, see
Fig.~\ref{et-u=1.5-n=1.0-etabare=0.2.eps}a. As one may expect, 
by shifting the 
 system away from its ground state, pairing  becomes energetically 
 less costly and for $n_{\vecQ}\lesssim 0.3$ it even lowers the energy. 
 Hence, the stiffness argument would predict a softening
 rather than a hardening of the excitations as shown in
the main panel of Fig.~\ref{del-u=1.5_n=1.0:etbare=0.2.eps}.

 One must 
 keep in mind that, while the expectation values of $\hat{n}_{\vecQ}$ and 
consequently of $\hat{H}_0$ become stationary in the long time limit, the 
 density matrix does not. Therefore, the stationary states considered in 
the left upper inset to  Fig.~\ref{del-u=1.5_n=1.0:etbare=0.2.eps} have not much in common 
 with the time-dependent long time states in 
Fig.~\ref{et-u=1.5-n=1.0-etabare=0.2.eps}, apart from the expectation value
of $\hat{n}_{\vecQ}$. It is therefore interesting to observe that
in an out-of-equilibrium situation the long-time limits of $\hat{n}_{\vecQ}$
yield the qualitatively correct behavior of the peak shifts in an
effective temperature ELRT  while the analysis of the stiffness in the
energy functional fails.



A charge-order probe pulse obviously detects the amplitude excitation
as shown in Fig.~\ref{et-u=1.5-n=1.0-etabare=0.2.eps}c.
This implies
 not only a shift of the peak position to lower energies
 but also a significant increase 
 of its weight. As already mentioned in the previous section the stark
 asymmetry in the spectra with respect to $\gamma=1$ is due to
 the fact that the underlying equilibrium CDW order parameter $\eta_{CO}$
 is close to its fully polarized value.

\begin{figure}[bt] 
\includegraphics[width=8cm]{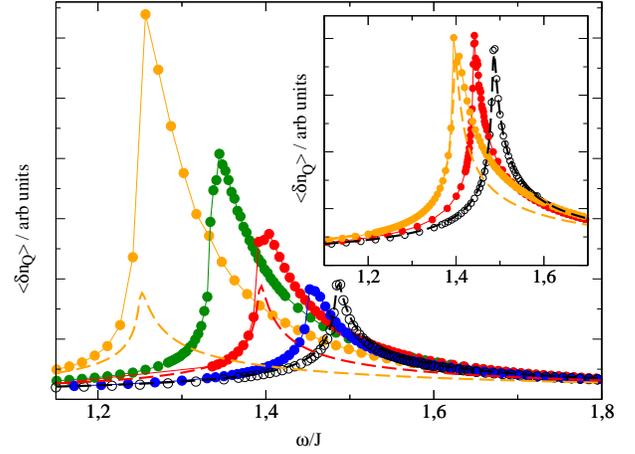} 
\caption{Charge-order response $\langle \delta n_{\vecQ}\rangle $
 as a function of frequency $\omega$
for $U/J=1.5$, $n=1.0$,  $\alpha_{\vecQ}=0.2$, 
and scaling factors $\gamma=1.0$ (black, dotted), 
$\gamma=0.6$ (blue), $\gamma=0.4$ (red), $\gamma=0.3$ (green), 
$\gamma=0.2$ (orange).
Inset: Same quantities for $\gamma=1.0$ (black, dotted), 
$\gamma=1.5$ (blue), $\gamma=2.0$ (red), $\gamma=2.5$ (green), 
$\gamma=3.0$ (orange). ELRT results are shown with dashed lines
and are obtained for the effective temperatures indicated
in the lower panel of Fig.~\ref{et-u=1.5-n=1.0-etabare=0.2.eps}.
}
\label{et-u=1.5_n=1.0:etbare=0.2.eps}
\end{figure}

We can also try to understand the peak positions from the ELRT analysis (cf. appendix
\ref{app3}). Without staggered field ($\alpha_Q=0$) the collective
CDW excitation appears at the frequency of the CDW gap $\omega=2\eta_{CO}$.
A finite $\alpha_Q$ pushes this excitation into the quasiparticle
continuum, however, the intensity of the RPA response function
$\chi^{\rm cdw}(\omega)=\chi_0^{\rm cdw}(\omega)/(1-V_Q \chi_0^{\rm cdw}(\omega))$
is still maximum at the CDW gap frequency due to (a) the enhancement
of the bare correlations $\chi_0^{\rm cdw}(\omega)$ and (b) the
minimum of $\Re\lbrack1-V_Q \chi_0^{\rm cdw}(\omega)\rbrack$ at
$\omega=2\eta_{CO}$. In fact, the peaks in the TDHF results shown
in Fig.~\ref{et-u=1.5_n=1.0:etbare=0.2.eps} occur exactly at the
values $\omega=2 \eta^\infty = |U|n_{\vecQ}^\infty + \alpha_{\vecQ}/2$
with $n_{\vecQ}^\infty$ being the long-time stationary value (cf.
Fig.~\ref{et-u=1.5-n=1.0-etabare=0.2.eps}a) from which
we have defined the effective temperatures and for which the
ELRT results are shown with the dashed curves in
Fig.~\ref{et-u=1.5_n=1.0:etbare=0.2.eps}.
Thus, while the peak positions between TDHF and effective temperature ELRT
show very good agreement, the intensities strongly deviate
in particular for $\gamma<1$ where the response in TDHF gets strongly
enhanced.

At first sight, one might think that this enhancement of spectral weight
 is a natural behavior that simply results from the larger amplitudes 
 of the underlying out-of-equilibrium oscillations 
in Fig.~\ref{et-u=1.5-n=1.0-etabare=0.2.eps}a. In fact, as we show in 
 Appendix~\ref{xcvv}, the very same behavior is found in the rather simple 
 model of
 a one-dimensional classical oscillator. 
It is crucial, however, to include
  an anharmonic term into the potential of that model. Without it, i.e., for a linear 
 equation of motion,  no  weight gain is 
 observed. We also do not find such a weight gain, when we calculate the exact 
out-of-equilibrium response function for a two-site (negative $U$) 
Hubbard model, see Appendix~\ref{app2}. This is not surprising because here, as well, 
we solve a set 
 of $4$ {\sl linear} differential equations with constant coefficients. 
 In contrast, the TDHF equations for the same model are non-linear. For this reason, we do observe a  
 weight gain which, at least for the two-site model, is clearly a spurious result. 
 We can 
therefore not rule out the possibility that the  weight-gain observations
 in the infinite 
 model that we presented in this section are artifacts of an oversimplifying 
  TDHF method. In contrast, our previous observation that position and 
weight of the
  pairing response function are rather robust when we go away from equilibrium 
 is confirmed by the exact results for the  two-site
Hubbard model in Appendix~\ref{app2}.

\subsubsection{Superconducting ground state}\label{xcv2}
Next we consider the case with
 $U/J=1.5$, $n=0.7$, $\alpha_{\vecQ}=0$
 where the system has a pure superconducting ground state.

\begin{figure}[bt]
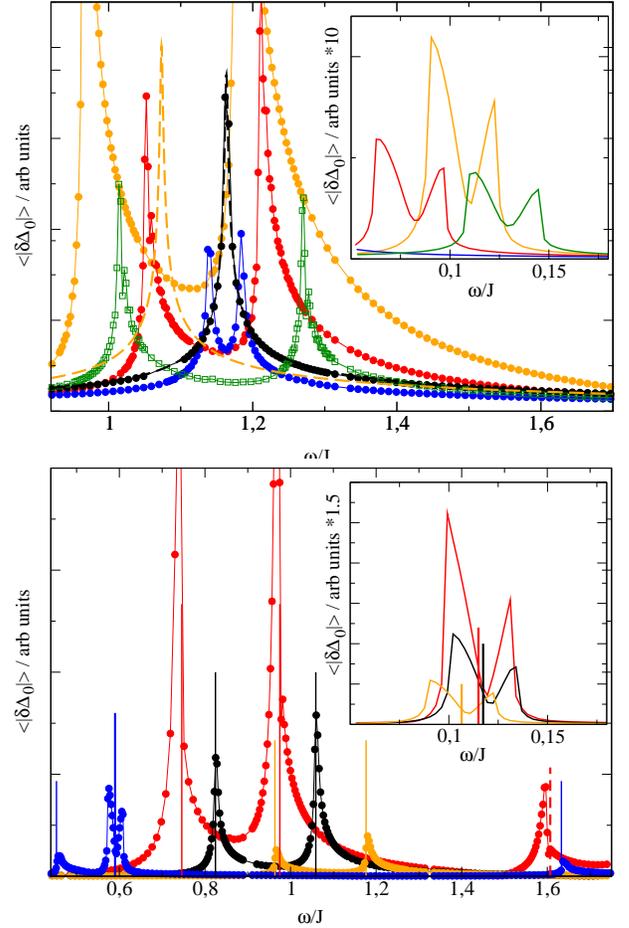

\includegraphics[width=8cm]{1.eps} \\
\includegraphics[width=8cm]{2.eps}
\caption
{Pairing response  $\langle \delta| \Delta_0|\rangle $
 as a function of frequency $\omega$
for $U/J=1.5$, $n=0.7$,  $\alpha_{\vecQ}=0$. Top:  
scaling factors $\gamma=1.0$ (black, dotted), 
$\gamma=0.9$ (blue), $\gamma=0.6$ (red), $\gamma=0.4$ (orange), 
$\gamma=1.5$ (green). Inset: same quantities for a smaller frequency
 range. Bottom: scaling factors $\gamma=0.4$ (orange), 
 $\gamma=0.2$ (black), $\gamma=0.125$ (red), $\gamma=3.0$ (blue).
 Inset: same quantities for a smaller frequency range. The vertical
 thin lines correspond to energies $\Omega_{A}\pm \Omega_{P}$ as deduced
 from the Fourier transform of amplitude and phase modes (cf.
 Fig.~\ref{del-u=1.5-n=0.7-etabare=0.0.eps}). The thick lines are phase
 modes whereas the red dashed vertical line is an excitation at
 $2\Omega_{A}- \Omega_{P}$ for $\gamma=0.125$.
}
\label{del-u=1.5_n=0.7:etbare=0.0.eps}
\end{figure}

We start our discussion with the pairing response 
function which is shown in Fig.~\ref{del-u=1.5_n=0.7:etbare=0.0.eps}
for scaling factors $\gamma$ between $0.125$ and $3$ for which the
corresponding amplitude and phase excitations have been reported
in Fig.~\ref{del-u=1.5-n=0.7-etabare=0.0.eps}c.

In the equilibrium limit ($\gamma=1$) the TDHF response is perfectly
reproduced by ELRT (dashed) which describes the amplitude excitation
(or `Higgs mode') at twice the superconducting gap. Upon increasing the
non-equilibrium situation (i.e., deviating from $\gamma=1$) one observes
two main features beyond the ELRT expectation: first, the
amplitude excitation splits into two peaks and second, additional weight
is observed at low energies. In fact, applying the effective temperature ELRT
yields a single peak located between the excitations of the TDHF result
(cf. result for $\gamma=0.4$ in the upper panel of Fig.~\ref{del-u=1.5_n=0.7:etbare=0.0.eps}). 
The reason for the splitting of the amplitude excitation
has been discussed in Sec. \ref{qyx} and is related to the definition
of the pairing response function Eq.~(\ref{rtz2}) which for the present situation
is influenced by the phase mode of the underlying
non-equilibrium state
(cf. Fig.~\ref{del-u=1.5-n=0.7-etabare=0.0.eps}c.) Therefore the response
$\langle|\delta\Delta_0|\rangle$ is determined by excitations appearing at
frequencies $\Omega_{A}\pm \Omega_{P}$, where $\Omega_{A,P}$ correspond to amplitude 
and phase modes, respectively.
The lower panel of Fig.~\ref{del-u=1.5_n=0.7:etbare=0.0.eps} demonstrates
the consistency of our analysis for various $\gamma$ values.

In contrast, the effective temperature ELRT only yields a renormalized (softened) 
amplitude mode and thus cannot account for the splitting when the response
is evaluated within Eq.~(\ref{10.1301}). Also an evaluation based on
Eq.~(\ref{10.1300}) does not yield a splitting since the phase mode
in ELRT always occurs at $\Omega_P=0$.

Moreover, for strong non-equilibrium initial states
 one observes further non-linear
processes as the appearance of an excitation at $2\Omega_{A}- \Omega_{P}$
for $\gamma=0.125$. As mentioned above the second feature concerns the
low energy spectral weight as shown in the insets to Fig.~\ref{del-u=1.5_n=0.7:etbare=0.0.eps} 
which can also be attributed to the coupling between phase
and amplitude degrees of freedom at strong non-equilibrium.
These low energy excitations are also split by $\omega_P\pm g^2/\Omega_A$ where $g$ denotes 
an effective coupling parameter between amplitude and phase.

\begin{figure}[bt] 
\includegraphics[width=8cm]{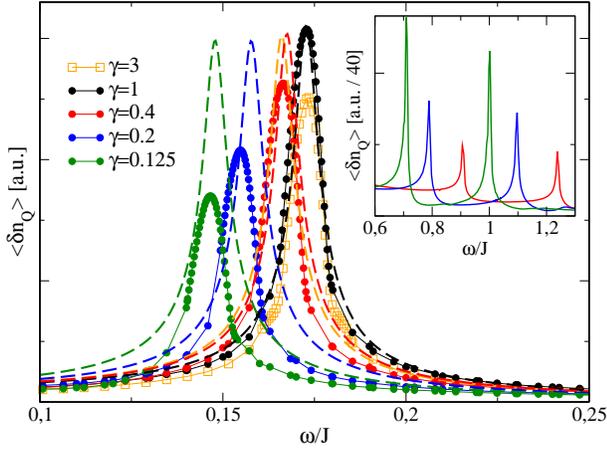} 
\caption{Charge-order response 
 $ \langle \delta n_{\vecQ}\rangle $
 as a function of frequency $\omega$ for $U/J=1.5$, $n=0.7$, $\alpha_{\vecQ}=0$, and scaling factors $\gamma=3.0$ (black, solid), $\gamma=1.0$ (black, dotted), 
$\gamma=0.4$ (red), $\gamma=0.2$ (blue), $\gamma=0.125$ (green): 
Inset: Same quantities at larger frequencies.}
\label{et-u=1.5_n=0.7:etbare=0.0.eps}
\end{figure}

 We proceed with the analysis of the charge-order response which
 is shown in Fig.~\ref{et-u=1.5_n=0.7:etbare=0.0.eps} again
 for various initial non-equilibrium situations parametrized by $\gamma$.
At half-filling the charge response at ${\bf q}=\vecQ$ would occur at  
zero frequency due to the degeneracy between CDW and SC. Finite doping
shifts this excitation to finite frequency inside the SC gap
(cf. Fig.~1 in Ref.~[\onlinecite{cea15}]) and
Fig.~\ref{et-u=1.5_n=0.7:etbare=0.0.eps} reveals that in the
equilibrium situation ($\gamma=1$) our TDHF response is perfectly described
by ELRT (dashed). For stronger non-equilibrium situations the effective
temperature ELRT still gives a reasonable description of the mode softening
but does not capture the reduction in intensity of the TDHF charge-order
response. Moreover, in non-equilibrium the charge order excitations at
${\bf q}=\vecQ$ can
couple to the pairing modes at ${\bf q}=0$ (cf. Fig.~\ref{del-u=1.5_n=0.7:etbare=0.0.eps} which induces the high energy features shown in the inset to
Fig.~\ref{et-u=1.5_n=0.7:etbare=0.0.eps}. Note that in this case the
splitting is not exactly $2\Omega_P$ but also influenced by the coupling
strength between charge and pairing modes.
 
\subsubsection{Ground state with finite pairing and charge order parameters}\label{xcv3}
Finally, as an example for a coexistence phase we consider the 
 the case where  $U/J=1.5$, $n=0.8$, $\alpha_{\vecQ}=0.2$. For these 
 parameters, $\Delta_0$ and $n_{\vecQ}$ have approximately the same values (cf.  Fig.~\ref{dn-del-of_U.eps}) and
 none of the two orders dominates.

\begin{figure}[bt] 
\includegraphics[width=8cm]{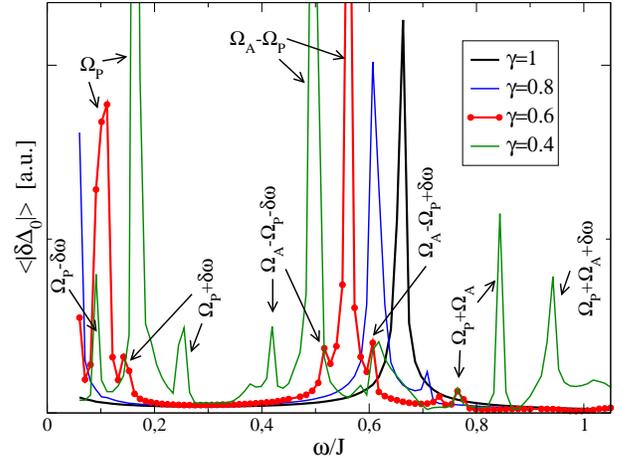} 
\caption{Pairing response  $\langle \delta| \Delta_0|\rangle $
 as a function of frequency $\omega$
for $U/J=1.5$, $n=0.8$,  $\alpha_{\vecQ}=0.2$
and scaling factors $\gamma=1.0$ (black, dotted), 
$\gamma=0.8$ (blue), $\gamma=0.6$ (red), $\gamma=0.4$ (green).
Coupled amplitude and phase excitations as deduced from
Fig.~\ref{ampphascdw+sc} are indicated for the $\gamma=0.4$, $0.6$ results. }
\label{del-u=1.5_n=0.8:etbare=0.2.eps}
\end{figure}

In Fig.~\ref{del-u=1.5_n=0.8:etbare=0.2.eps} we show the pairing 
response in the equilibrium ($\gamma=1$) and non-equilibrium situation
($\gamma\ne 0$). The linear-response for a coupled CDW-SC system has
been recently analyzed in Ref. \onlinecite{cea142} in the context of
the visibility of the amplitude ('Higgs') mode within 
linear (Raman) response. \cite{cea142}. In fact, the
presence of CDW order pushes the linear response amplitude excitation
to $\Omega_A\approx 0.66 J$, i.e. well below
the $-E_-(\veck_F) \to +E_-(\veck_F)\approx 1.2 J$
transition [cf. Eq.~(\ref{eq:evals})] where it would appear in the pure SC
case. 
In the non-equilibrium case the individual peaks can
be understood from an inspection of the corresponding time
evolution of the order parameters as shown in Fig.~\ref{ampphascdw+sc}
for the case $\gamma=0.6$ and from the general structure of the
pairing response function as given in Eq.~(\ref{10.1300}).
The latter couples the phase to the amplitude modes, similar to what
is already observed in the bare SC case Fig.~\ref{del-u=1.5_n=0.7:etbare=0.0.eps}, yielding excitations at $\Omega_A\pm\Omega_P$ plus the phase mode
at $\Omega_P$. An additional feature concerns the interference effect
between phase and amplitude excitation which occurs at
\begin{equation}
\delta\omega=\Omega_A-6\Omega_P
\end{equation}
 and generates further satellite peaks
to the main excitations discussed above.

\begin{figure}[bt] 
\includegraphics[width=8cm]{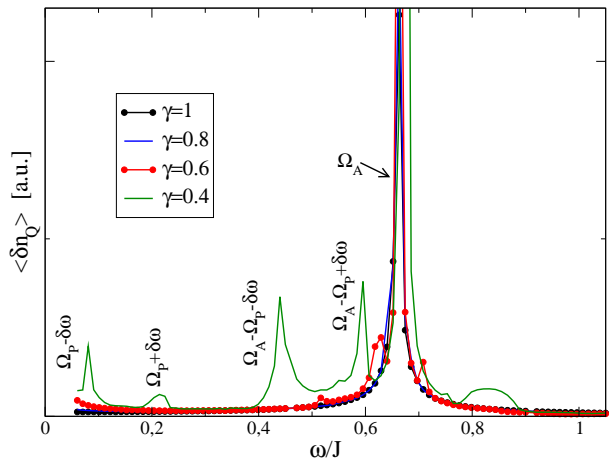} 
\caption{Charge order response 
 $\langle \delta n_{\vecQ}\rangle $
 as a function of frequency $\omega$
for $U/J=1.5$, $n=0.8$,  $\alpha_{\vecQ}=0.2$
and scaling factors $\gamma=1.0$ (black, dotted), 
$\gamma=0.8$ (blue), $\gamma=0.6$ (red), $\gamma=0.4$ (green).
Coupled amplitude and phase excitations as deduced from
Fig.~\ref{ampphascdw+sc} are indicated for the $\gamma=0.4$, $0.6$ results.}
\label{et-u=1.5_n=0.8:etbare=0.2}
\end{figure}

Since we consider a commensurate charge order at ${\bf Q}=(\pi,\pi)$
the corresponding order parameter can always be chosen as real and
there is no associated phase degree of freedom. Thus the
charge order response, as shown in Fig.~\ref{et-u=1.5_n=0.8:etbare=0.2}
occurs at $\Omega_A$. Note, however, that the charge and SC amplitudes
are coupled which induces the satellite peaks at $\Omega_A\pm \delta\omega$.
Increasing the non-equilibrium situation to $\gamma=0.4$ induces
further satellites related to phase modes and their coupling to
the amplitude excitations.
In any case, it is obvious that in the non-equilibrium situation both,
charge and pairing response cannot be captured by ELRT which
does not account neither for the coupling of amplitude and phase nor
the satellite structure due to the interference scale $\delta\omega$.

\section{Conclusions}\label{conclusions}
In this work, we have investigated
 pairing and charge-order response functions 
 in out-of-equilibrium states of the negative $U$ Hubbard model by means 
 of the time-dependent Hartree-Fock approximation (TDHF). In particular,
 we have focussed on the coupling between amplitude and phase
 excitations which can be inherent in the definition of the
 response function [cf. Eq.~(\ref{eq:d0c})] but also be induced
 in a non-equilibrium situation.

 We have allowed for an 
 additional charge-order field in the Hamiltonian which simulates the 
 effect of a lattice modulation. In this way, our 
 model can have three types of ground states, a pure superconducting state, 
 a charge-order state, or a state with both orders present. A pump-pulse 
 may then drive the system into non-equilibrium states of the same symmetry.
  
Since the TDHF can be applied in all non-equilibrium situations, we did not 
have to rely on any linear-response assumptions that are inherent, e.g., in 
 the Kubo formula. In this way our study also revealed, if and under what 
circumstances such an assumption is justified in  non-equilibrium 
calculations. It turned out that a linear-response assumption,
based on an effective temperature description of the
underlying non-equilibrium state, qualitatively accounts for the
spectra when the latter is due to a pure CDW dynamics without
any coupling to phase degrees of freedom. In these cases 
(Figs.~\ref{del-u=1.5_n=1.0:etbare=0.2.eps} and \ref{et-u=1.5_n=1.0:etbare=0.2.eps}) ELRT
works for not too large deviations from $\gamma=1$ but fails
to reproduce the excitation energy or (and) peak intensity
in stronger non-equilibrium situations.
For an underlying non-equilibrium SC state the appearance
of a finite frequency phase mode has significant consequences
for the pairing and charge order response. Concerning the
latter, ELRT correctly describes the formation of in-gap
excitations (cf. Fig.~\ref{et-u=1.5_n=0.7:etbare=0.0.eps}), although
underestimating the intensity, while it fails do describe
the amplitude-phase coupling in strong non-equilibrium which
induces the appearance of split peaks on the scale of the SC gap.
The same holds in case of the pairing response where the splitting
is due to the appearance of the phase mode in the definition
of the response Eq.~(\ref{eq:d0c}) and when the non-equilibrium
state is an admixture of both, CDW and SC.

In this work, we have considered a neutral system without long-range Coulomb
interactions which due to the Anderson-Higgs mechanism \cite{anderson63}
would push the phase mode up to the plasma frequency. For a two-dimensional
system the plasma frequency would be still at
low energy $\omega_p \sim \sqrt{|{\bf q}|}$ so that for these systems our results
could be still meaningful. Also disorder helps to push the plasma frequency
to lower energies due to the reduced superfluid density.

The interplay of CDW and SC in the attractive Hubbard model
has been investigated previously \cite{cea142} in the context of the visibility
of the amplitude (`Higgs') mode in charge-density wave superconductors
such as NbSe$_2$. These authors were interested in the Raman response
which amounts to the evaluation of the charge response function at
momentum ${\bf q}=0$. Here, instead we have studied the charge response function
at the momentum of the CDW ${\bf q}={\bf Q}$ which in principle can
be measured with inelastic x-ray scattering or indirectly with neutrons
via the coupling to the lattice. The pairing response can be experimentally
accessed with the Josephson effect which has been previously used to
investigate the contribution of pair fluctuations to the pseudogap
formation in high-T$_c$ superconductors. \cite{bergeal08}

While the measurement of these responses for CDW superconductors
would be definitely interesting to compare with our predictions,
an equally important issue of the present paper concerns the
validity of linear-response theory in a non-equilibrium situation
with regard to pump-probe experiments. Methodologically our investigations
are based on the TDHF approximation which can be viewed as
the simplest approach to study the dynamics of interacting
systems and therefore one has to be aware of its limitations.
In this regard, one aspect concerns the damping of the order parameter
which in TDHF is caused by a `dephasing' of oscillations for the 
different Hartree Fock single-particle energies. Genuine many-particle relaxation
processes are not covered by the TDHF and it is therefore bound to
become inaccurate in the 
limit $t\gg J^{-1}$. In pump-and-probe experiments, however, the probe pulse
is usually applied in a time period where the excitation induced by the
pump pulse is still far from relaxation. Hence, we are confident, that the
TDHF constitutes a 
 meaningful first-order approximation to these problems.

We have also critically 
examined our observations by looking into two simple toy models.  
 Our findings seem to be confirmed  
when we consider a simple anharmonic classical oscillator. Due to the 
  non-linear term in the equation of motion, the response to a small 
  external field also depends strongly on the amplitude of the  underlying 
 oscillation. In contrast, however, the exact solution for the charge-order 
response  of a 
 two-site Hubbard model is, in this regard, different from what the  
 TDHF method finds. The reason for this difference
  is the non-linearity that the  TDHF spuriously
 introduces into its equations of motion.  
It remains an open question, if these 
 deficiencies of the TDHF are a generic problem of that method or can be 
 explained by the particular nature of the low-dimensional two-site Hubbard 
model. 

For the negative $U$ Hubbard model, there are more sophisticated methods 
 available that could be used to study the out-of-equilibrium response functions
 which we have investigated in this work. The most obvious way to improve 
 the TDHF is to use 
 Gutzwiller wave functions instead of single-particle product wave 
functions.~\cite{schiro2010,schiro2011,andre2012,fabrizio2013,sandri2013,buenemann2013a,buenemann2013b,buenemann2015a}
Work in this direction is in progress. 
  Other conceivable approaches are the out-of-equilibrium DMFT or purely numerical 
methods such as DMRG, quantum Monte-Carlo, or exact diagonalization. 
 Our TDHF results constitute a useful first-order approximation
 in all such future investigations. 

\section*{Acknowledgements}
This work was supported by  the Deutsche Forschungsgemeinschaft (DFG)
under SE 806/18-1.

\appendix

\section{Minimisation algorithm}
\label{app1}
 For the numerical minimization of our energy functional, we need to
 solve the equation
\begin{equation}\label{iopj}
[\tilde{h}_{\varepsilon_i}, \tilde{\rho}_{\varepsilon_i}]=0
\end{equation}
self-consistently for all energies $\varepsilon_i$ and 
 obeying the additional 
 constraint  $\tilde{\rho}^2_{\varepsilon_i}=\tilde{\rho}_{\varepsilon_i}$
 for single-particle wave functions. This equation is readily solved numerically
 by determining the eigenvectors and eigenvalues of $\tilde{h}_{\varepsilon_i}$. 
 We have determined the minimum with the following 
 algorithm
\begin{itemize}
\item[i)] We start with some small, but non-zero, (input) values for the fields 
$\eta^{\rm i}_{\rm sc}, \eta^{\rm i}_{\rm co}, \delta \eta^{\rm i}_{\rm sc}$ and set up the matrices
 $\tilde{h}_{\varepsilon_i}$ for each of the $N_{\rm disk}$ energies 
$\varepsilon_i$. 
\item[ii)] For each $\varepsilon_i$,  we solve Eq.~(\ref{iopj}).
\item[iii)]
With  $\tilde{\rho}_{\varepsilon_i}$, new values
 $\eta^{\rm o}_{\rm sc}, \eta^{\rm o}_{\rm co}, \delta \eta^{\rm o}_{\rm sc}$ are 
determined
 using (\ref{yxc})-(\ref{yxc2}) and (\ref{yvb1})-(\ref{yvb2}).
With these values, we could go back to i). However, it is usually necessary
to introduce some `damping factor' $\beta<1$ and continue with, e.g., 
$ \eta^{\rm i}_{\rm sc}\to  \eta^{\rm i}_{\rm sc}+\beta (\eta^{\rm o}_{\rm sc}
-\eta^{\rm i}_{\rm sc})$.
Without such a damping, it is not ensured that the energy  decreases in each 
step of our algorithm.
 \item[iv)]The algorithm terminates when the fields in iii) are approximately 
 the same as the input values in i). 
\end{itemize}

\section{Equilibrium linear-response theory (ELRT) for a SC perturbation of a CDW ground state}
\label{app3} 

The Hamiltonian for a CDW is given by
\begin{equation}\label{hamcdw} 
  H=\sum_{\veck,\sigma}(\varepsilon_{\veck}-\mu)\hat{c}_{\veck,\sigma}^\dagger 
\hat{c}_{\veck,\sigma}
  +\eta \sum_{\veck,\sigma}(\varepsilon_{\veck}-\mu)
\hat{c}_{\veck+\vecQ,\sigma}^\dagger \hat{c}_{\veck,\sigma}
\end{equation}
with $\vecQ=(\pi,\pi)$ and $\varepsilon_{\veck}=-\varepsilon_{\veck+\vecQ}$.
The transformation
\begin{eqnarray*}
  \hat{c}_{\veck,\sigma}&=& \beta_{\veck} 
\hat{a}_{\veck,-,\sigma} +\alpha_{\veck} \hat{a}_{\veck,+,\sigma} \;,\\
  \hat{c}_{\veck+\vecQ,\sigma}&=& -\alpha_{\veck} \hat{a}_{\veck,-,\sigma} 
+\beta_{\veck} \hat{a}_{\veck,+,\sigma}
\end{eqnarray*}
diagonalizes Eq.~(\ref{hamcdw}) and yields
\begin{equation} 
  H=\sum_{\veck\in{\mathcal B}_0,\sigma}\left[(-\mu+E^+_{\veck})
\hat{a}_{\veck,+,\sigma}^\dagger \hat{a}_{\veck,+,\sigma}
    +(-\mu+E^-_{\veck})\hat{a}_{\veck,-,\sigma}^\dagger 
\hat{a}_{\veck,-,\sigma}\right]
\end{equation}
with
\begin{eqnarray*}
  \alpha_{\veck} &=& \frac{1}{\sqrt{2}} \sqrt{1+\frac{\varepsilon_{\veck}}{E_{\veck}}}\;, \\
  \beta_{\veck} &=& \frac{1}{\sqrt{2}} \sqrt{1-\frac{\varepsilon_{\veck}}{E_{\veck}}} 
\end{eqnarray*}
where $E_{\veck}^{\pm}=\pm \sqrt{\varepsilon_{\veck}^2+\Delta^2}$ and ${\mathcal B}_0$ 
is the reduced Brillouin zone introduced in Sec.~\ref{ham}.

\subsection{Pair fluctuations}
We use the pairing operators $\hat{\Delta}_0$ and $\hat{\Delta}_{\vecQ}$, 
as introduced in (\ref{yxc}) and
 (\ref{yxcb}) 
to define the corresponding pair correlation functions
\begin{eqnarray*}
\chi_0^{\Delta_0^\dagger\Delta_0}&=&-i\langle {\cal T} \hat{\Delta}_0^\dagger
\hat{\Delta}_0 \rangle\;,\\
\chi_0^{\Delta_\vecQ^\dagger\Delta_\vecQ}&=&-i\langle {\cal T} \hat{\Delta}_\vecQ^\dagger
\hat{\Delta}_\vecQ \rangle\;,\\
\chi_0^{\Delta_0^\dagger\Delta_\vecQ}&=&-i\langle {\cal T} \hat{\Delta}_0^\dagger
\hat{\Delta}_\vecQ \rangle\;,\\
\chi_0^{\Delta_\vecQ^\dagger\Delta_0}&=&-i\langle {\cal T} \hat{\Delta}_\vecQ^\dagger
\hat{\Delta}_0 \rangle \;.
\end{eqnarray*}
One obtains
\begin{eqnarray*}
 && \chi_0^{\Delta_0^\dagger\Delta_0}(\omega)=-\frac{1}{N}\sum_{\veck\in{\mathcal B}_0 ,s=\pm}
  \frac{1-2f(E_{\veck}^s - \mu)}{\omega-2\mu+2E_{\veck}^s}\;, \\
&&  \chi_0^{\Delta_\vecQ^\dagger\Delta_\vecQ}(\omega)=-\frac{1}{N}\sum_{\veck ,s=\pm}
\frac{\eta^2}{E_{\veck}^2}  \frac{1-2f(E_{\veck}^s - \mu)}{\omega-2\mu+2E_{\veck}^s}\;, \\
  &&-\frac{1}{N}\sum_{\veck\in{\mathcal B}_0 ,s=\pm}
\frac{\varepsilon_{\veck}^2}{E_{\veck}^2}  \frac{1-f(E_{\veck}^+ - \mu)-f(E_{\veck}^- - \mu)}{\omega-2\mu}\;,\\
&&\chi_0^{\Delta_\vecQ^\dagger\Delta_0}(\omega)=\chi_0^{\Delta_0^\dagger\Delta_\vecQ}(\omega)\;,\\
&&=-\frac{1}{N}\sum_{\veck\in{\mathcal B}_0 ,s=\pm}s
  \frac{\eta}{E_{\veck}}\frac{1-2f(E_{\veck}^s - \mu)}{\omega-2\mu+2E_{\veck}^s}\;,
\end{eqnarray*}
which we evaluate with the density of states $D(\varepsilon)$
as defined in Eq.~(\ref{dep}).

We now consider a perturbation from pairing fluctuations as, e.g., arising
from an attractive on-site interaction
\begin{equation}\label{scperturb}
   \hat{V}=-\frac{|U|}{N}\sum_{\bf q} \hat{\Delta}_{\bf q}^\dagger 
\hat{\Delta}_{\bf q}
\end{equation}
with 
\begin{equation}\label{scperturb1}
 \hat{\Delta}_{\bf q}\equiv 
\frac{1}{L}\sum_{\veck} \hat{c}_{\veck,\uparrow}^{\dagger}
 \hat{c}_{-k-{\bf q},\downarrow}^{\dagger}\;,
\end{equation}
which yields the following RPA problem for the pair correlation
functions
\begin{equation}
  \underline{\underline{\chi}}=\underline{\underline{\chi_0}}
  +\underline{\underline{\chi_0}}\underline{\underline{V}}\underline{\underline{\chi}}
\end{equation}
with
\begin{equation}
  \underline{\underline{\chi_0}}=
  \left(\begin{array}{cc}
    \chi_0^{\Delta_0^\dagger\Delta_0} & \chi_0^{\Delta_0^\dagger\Delta_\vecQ} \\
    \chi_0^{\Delta_\vecQ^\dagger\Delta_0} & \chi_0^{\Delta_\vecQ^\dagger\Delta_\vecQ}
    \end{array}\right)
\end{equation}
and
\begin{equation}
  \underline{\underline{V}}=
  \left(\begin{array}{cc}
    -|U| & 0 \\
    0 & -|U|
    \end{array}\right)  \,.
\end{equation}

\begin{figure}[bt] 
\includegraphics[width=7.5cm]{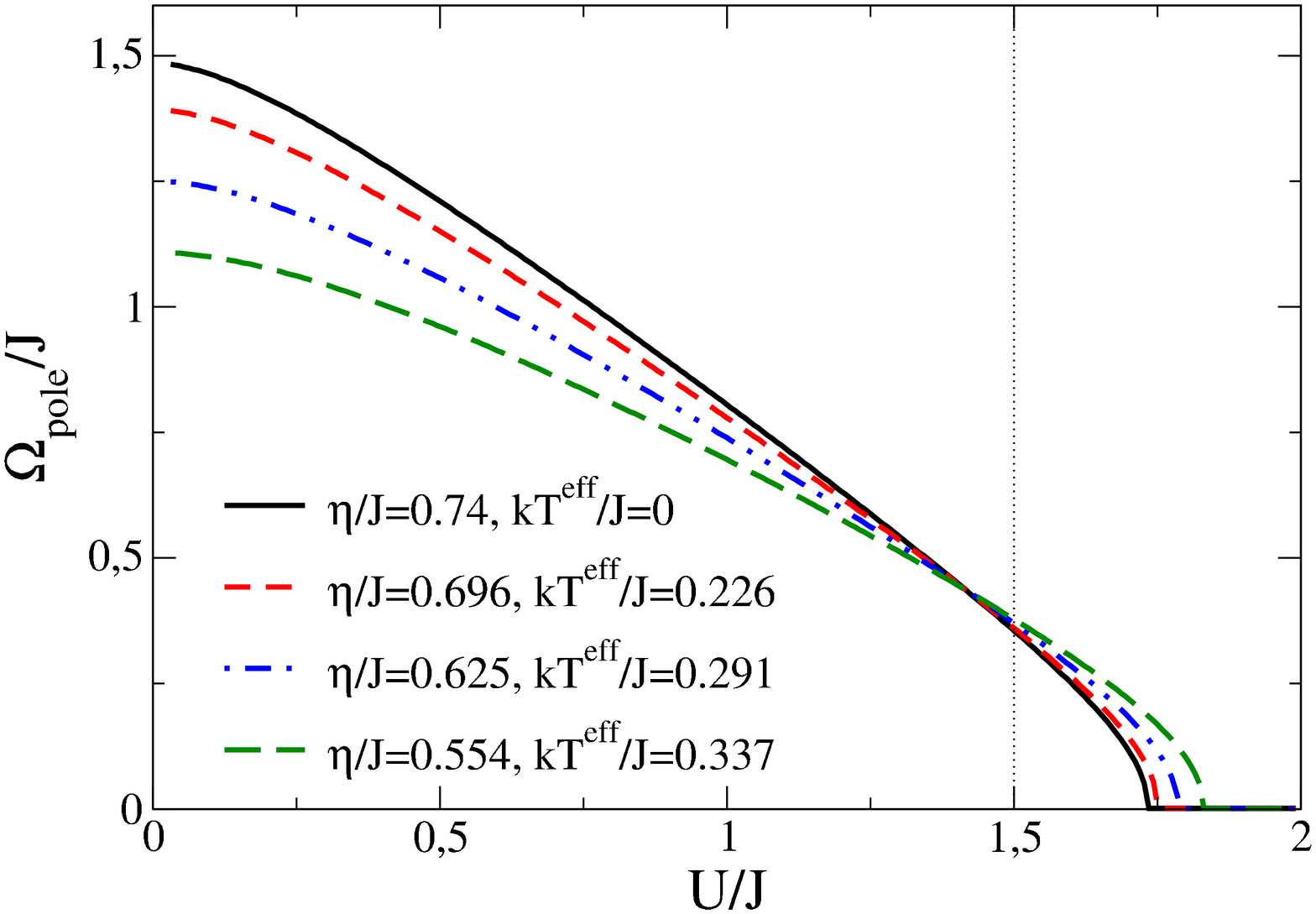} 
\caption{Poles $\Omega_{\rm pole}$ within the CDW gap $2\eta$
  induced by a SC perturbation Eq.~(\ref{scperturb}) at half-filling
  and for
  the temperatures for which the ELRT results are shown in
  Fig.~\ref{del-u=1.5_n=1.0:etbare=0.2.eps}.
}
\label{poles}
\end{figure}

The poles $\Omega_{\rm pole}$ within the CDW gap $2\eta$ can then
be determined from the condition
\begin{equation}\label{eq:det}
  DET |\underline{\underline{1}}-  \underline{\underline{\chi_0}}  \underline{\underline{V}}|=0
\end{equation}
which for $\omega=0$ corresponds to the standard
Thouless criterion for a SC instability. The solutions of Eq.~(\ref{eq:det})
are shown  in Fig.~\ref{poles} and with increasing $U/J$ move from
the CDW gap at $\Omega_{\rm pole}=2\eta_{CO}$ to $\Omega_{\rm pole}=0$
where the SC instability
is reached. Note, however,  that the curves in Fig.~\ref{poles} are
obtained for {\it fixed} $\eta$ whereas in the present HF theory 
$\eta_{CO}$ itself is an increasing function of $|U|$
so that the instability is never reached at half-filling.

\subsection{CDW fluctuations}
The (Hermitian) operator $\hat{n}_{\vecQ}$ for CDW fluctuations has been defined 
in~(\ref{yxc2})
and the corresponding CDW correlation function reads
\begin{displaymath}
\chi_0^{
\rm cdw}=-i\langle {\cal T} \hat{n}_{\vecQ}
\hat{n}_{\vecQ} \rangle 
\end{displaymath}
which can be evaluated as
\begin{displaymath}
  \chi_0^{
\rm cdw}(\omega)= \frac{4}{N}\sum_{\veck\in{\mathcal B}_0,\sigma}\frac{\varepsilon_{\veck}^2}{E_{\veck}}
  \frac{f(E_{\veck}^-)-f(E_{\veck}^+)}{\omega^2-4 E_{\veck}^2}\,.
\end{displaymath}

The interaction between the CDW fluctuations is given by
$V_\vecQ=1/2 (\frac{-|U|}{2})\delta \Delta_\vecQ\delta \Delta_{-\vecQ}$
so that the RPA result for the correlation function is obtained
as
\begin{equation}\label{chicdw}
\chi^{
\rm cdw}(\omega)=\frac{\chi_0^{
\rm cdw}(\omega)}{1-V_\vecQ \chi_0^{
\rm cdw}(\omega)}\,.
\end{equation}
In particular, for $\omega=2\eta$, i.e., at the energy of the CDW
gap the denominator of Eq.~(\ref{chicdw})
\begin{equation}\label{eq:chi2delta}
1-V_\vecQ \chi_0^{
\rm cdw}(\omega)=1-\frac{|U|}{2N}\sum_{\veck\in{\mathcal B}_0,\sigma}
\frac{f(E_{\veck}^-)-f(E_{\veck}^+)}{E_{\veck}}
\end{equation}
vanishes when the external staggered field $\alpha_\vecQ=0$.
In fact, in this case Eq.~(\ref{eq:chi2delta}) is identical to
the self-consistency equation for the CDW order parameter $\eta_{CO}$,
so that for $\alpha_\vecQ=0$ the CDW amplitude excitations occur exactly
at $\omega=2\eta_{CO}$ and therefore are damped due to their
admixture with the quasiparticle excitations. With finite
(positive) $\alpha_\vecQ$ the amplitude excitation is further pushed
into the continuum, however, since the real part of
$1-V_\vecQ \chi_0^{
\rm cdw}(\omega)$  still acquires a minimum at $\omega=2\eta_{CO}$
the linear-response spectra in Fig.~\ref{et-u=1.5_n=1.0:etbare=0.2.eps}
are peaked at the energy of the CDW gap.

\section{Anharmonic oscillator: a classical example for 
out-of-equilibrium response functions}\label{xcvv}
As a simple illustrative example we show results for
 a one-dimensional classical 
 (anharmonic) oscillator. It is described by the differential equation
\begin{equation}\label{xtz}
\ddot{x}=-x+\alpha x^3+\Omega \sin(\omega t)
\end{equation}
with ($\Omega\neq 0$) or without ($\Omega= 0$) an external 
 frequency-dependent perturbation. We solve~(\ref{xtz}) 
numerically in a time interval 
$0\le t\le \Delta t $ with the initial condition 
   $x(0)=x_0$ and   $\dot{x}(0)=0$. The solution 
with and without external perturbation are denoted as  
$x_{\omega}(t)$ and 
 $x_0(t)$, 
respectively. Like in the main part of this work, we 
define the time-dependent deviation
\begin{equation}
\delta x(t)=x_{\omega}(t)-x_0(t) 
\end{equation}
and its expectation value
\begin{equation}\label{asd}
\langle \delta x\rangle =\frac{1}{\Delta t}\int_0^{\Delta t} {\rm d}t  
|x_{\omega}(t)-x_0(t)  |\;,  
\end{equation}
which is a function of the external frequency $\omega$.
The expectation value~(\ref{asd}) serves as a  measure of the system's response to  the external
 perturbation. 

\begin{figure}[bt] 
\includegraphics[width=8cm]{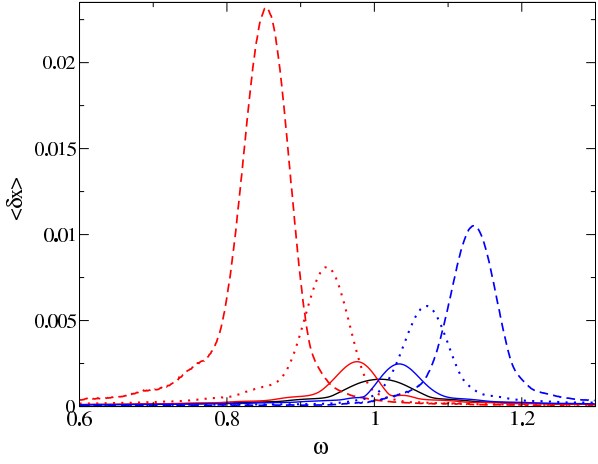} 
\caption{Response function $\langle  \delta x  \rangle$ of the anharmonic 
classical oscillator as a function of frequency $\omega$ 
for $\Omega=10^{-4}$, $\Delta t=100 $,  $\alpha=1$ (red curves), 
$\alpha=-1$ (blue) and initial amplitudes
 $x_0=0.2$ (solid),  $x_0=0.4$ (dotted), $x_0=0.6$ (dashed). For $x_0=0$ 
 the results for  $\alpha=1$ and  $\alpha=-1$ are the same (black curve).}
\label{oszz.eps}
\end{figure}

In Fig.~\ref{oszz.eps} we display $\langle \delta x\rangle$
 as a function of frequency around equilibrium ($x_0=0$) and for 
 several out-of-equilibrium amplitudes ($x_0\neq 0$). As expected, 
 the peak position is shifted towards higher (lower) frequencies when 
  $\alpha=-1$ ($\alpha=1$). Out of equilibrium, the weight of the 
 resonance grows substantially when the initial amplitude $x_0$ is increased.
   This is in agreement with our corresponding observations for 
 the negative $U$ Hubbard model in the time-dependent Hartree Fock
 approximation. Note that in the linear limit ($\alpha=0$)
 the response function $\langle \delta x\rangle$ is independent 
 of the amplitude $x_0$, i.e., it is the same at and away from 
 equilibrium. The frequency shifts and weight increases in 
 Fig.~\ref{oszz.eps} are therefore genuine effects of the 
 non-linear terms in the differential equation~(\ref{xtz}).

\section{Two-site Hubbard model: exact solution versus Hartree-Fock approximation}
\label{app2} 
The Hilbert space of a half filled two-site Hubbard model is four-dimensional
 when  we assume that the total spin $S_z$ in quantization direction is zero.
 A basis for this space may be chosen as 
  \begin{eqnarray}
    | d,0\rangle&=&\hat{c}^{\dagger}_{1,\uparrow} 
\hat{c}^{\dagger}_{1,\downarrow}|{\rm vac} \rangle\;, \\
 |0, d\rangle&=&\hat{c}^{\dagger}_{2,\uparrow} 
\hat{c}^{\dagger}_{2,\downarrow}|{\rm vac} \rangle\;. \\
 |\uparrow ,\downarrow\rangle&=&\hat{c}^{\dagger}_{1,\uparrow} 
\hat{c}^{\dagger}_{2,\downarrow}|{\rm vac} \rangle\;, \\
 | \downarrow ,\uparrow \rangle&=&\hat{c}^{\dagger}_{2,\uparrow} 
\hat{c}^{\dagger}_{1,\downarrow}|{\rm vac} \rangle\;.
 \end{eqnarray}
When we want to study a pairing probe pulse we further need to include
 the states $| 0,0\rangle=|{\rm vac} \rangle$ and
\begin{equation}
 | d,d\rangle=\hat{c}^{\dagger}_{1,\uparrow} \hat{c}^{\dagger}_{1,\downarrow}
  \hat{c}^{\dagger}_{2,\uparrow} \hat{c}^{\dagger}_{2,\downarrow} |{\rm vac} \rangle
\end{equation}
because the pairing operator $\hat{\Delta}_0$ has the form
\begin{equation}
  \hat{\Delta}_0=\frac{-1}{2}
\Big[
 | 0,0\rangle \big(  \langle 0,d|+   \langle d,0|\big)
+\big(  | 0,d\rangle +   | d,0\rangle\big)  \langle  d,d|\rangle
\Big]\;.
\end{equation}
The charge density-operator $n_{\vecQ}$ in the two-site model is given as
\begin{equation}
  \hat{n}_{\vecQ}=\frac{1}{2}
\big(  | 0,d\rangle   \langle 0,d|-
| d,0\rangle   \langle d,0|\big) \;.
\end{equation}
At half filling, the chemical potential is $\mu=-U/2$ and the operator 
 $\hat{K}$ hence becomes
\begin{eqnarray}\nonumber
\hat{K}&=&J\big(  
\langle d,0|+   \langle 0,d|
\big)
\big(  
\langle \uparrow,\downarrow|+   \langle \downarrow  ,  \uparrow |
\big)+{\rm h.c.}\\
&&+U  
\big(  
| \uparrow,\downarrow\rangle  \langle \uparrow,\downarrow    |
+| \downarrow,\uparrow\rangle  \langle \downarrow,\uparrow    |
\big)-\alpha_{\vecQ}\hat{n}_{\vecQ}\:.
\end{eqnarray}

\begin{figure}[t] 
\includegraphics[width=8cm]{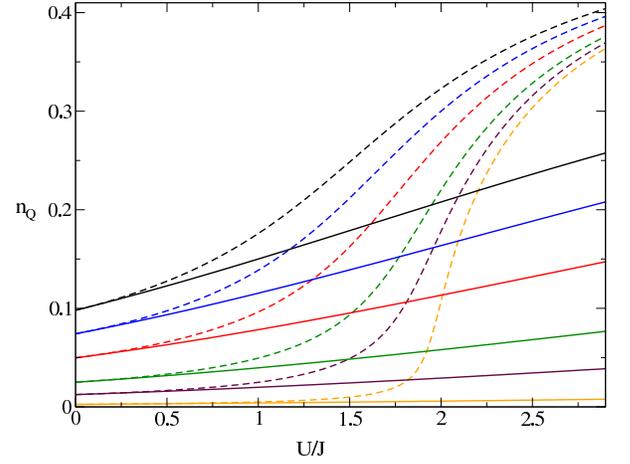} 
\caption{Exact (solid lines) and Hartree-Fock (dashed lines) 
 values for the charge density order $n_{\vecQ}$ in the ground state of the
 two site (negative $U$) 
Hubbard model as a function of $U/J$ and for $\alpha_{\vecQ}/J=0.4$ (black), 
 $0.3$ (blue), $0.2$ (red), $0.1$ (green), $0.05$ (maroon), $0.01$ (orange).
}
\label{nQ.eps}
\end{figure}

A Hartree-Fock approximation is particularly weak 
 for low-dimensional systems. Hence, one cannot expect it to  
 describe the physics of two-site Hubbard model satisfactorily. 
 The difficulties are already visible in the ground state 
 properties. In Fig.~\ref{nQ.eps} we show the exact and Hartree-Fock  
 values for the charge density order $n_{\vecQ}$  as a function of $U/J$ 
and for several values of $\alpha_{\vecQ}$. For $\alpha_{\vecQ}=0$ the 
 exact ground state shows no charge
 order whereas such an ordered state becomes stable in the  
Hartree Fock approximation
for $U/J>2$.   This is due to the fact that a `doublet singlet state' of the 
 form $| d,0\rangle+| 0,d\rangle$ becomes the ground state for $U\gg J$ and 
cannot be described within the Hartree Fock approximation. For  
$\alpha_{\vecQ}\neq 0$ the exact and Hartree Fock results in Fig.~\ref{nQ.eps} 
differ only 
 quantitatively, however, these differences are 
 quite substantial in large parts of the parameter space.
\begin{figure}[t] 
\includegraphics[width=8cm]{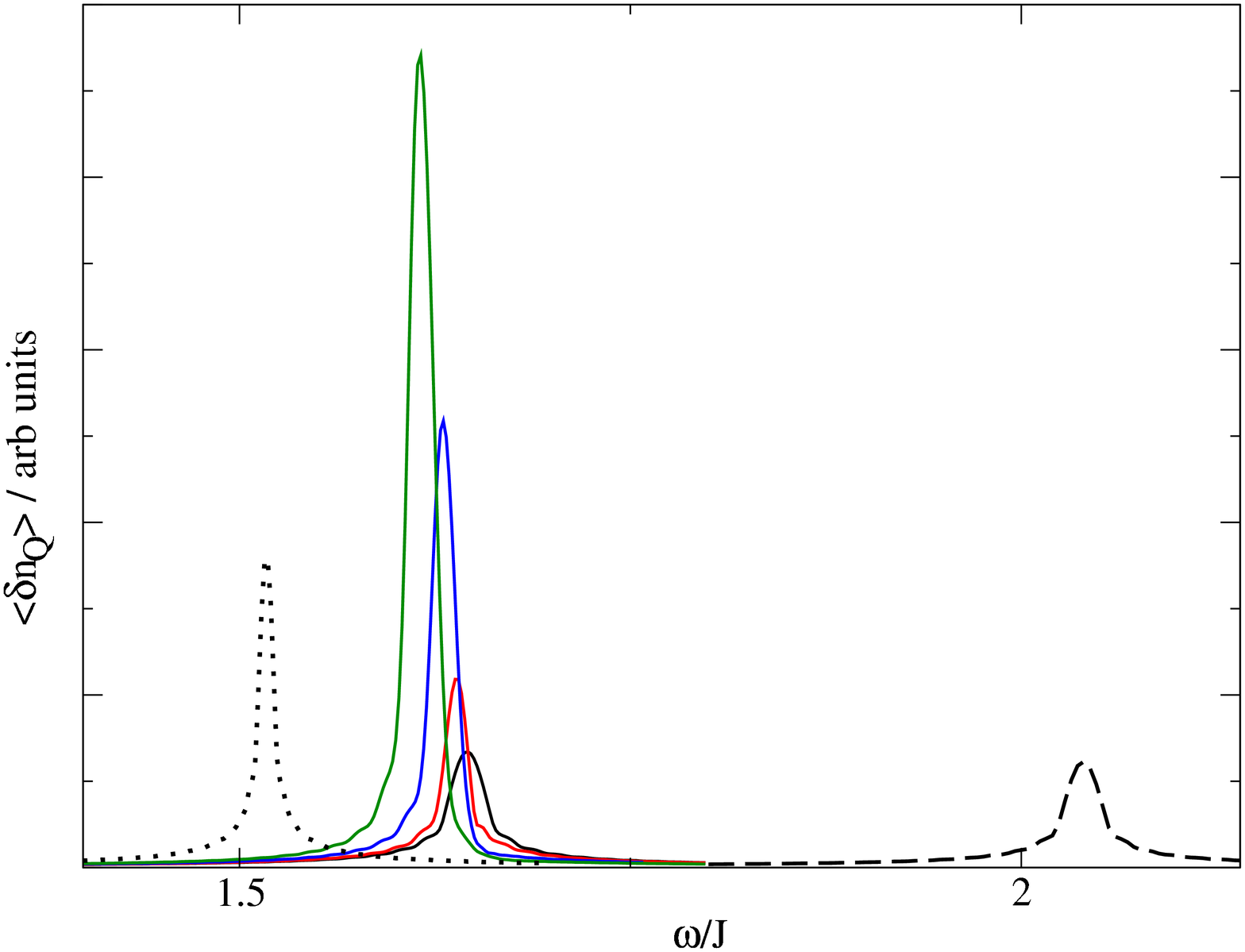} 
\caption{Charge-order response $\langle \delta n_{\vecQ}\rangle $ 
 as a function of frequency
 for $U/J=1.5$ (dotted for exact and solid for TDHF results) 
and $U/J=0.0$ (dashed) with  scaling factors
 $\gamma=1.0$ (black), $\gamma=0.8$ (red),  $\gamma=0.6$ (blue), and  
$\gamma=0.4$ (blue).
}
\label{pl10.eps}
\end{figure}

\begin{figure}[b] 
\includegraphics[width=8cm]{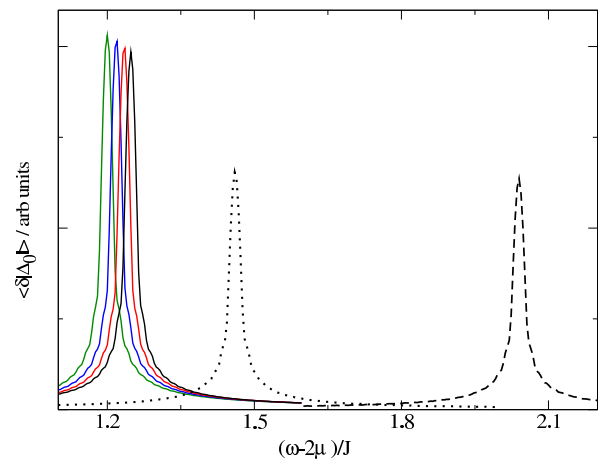} 
\caption{Pairing response  $\langle \delta| \Delta_0|\rangle $ as a function of frequency
 for $U/J=1.5$ (dotted for exact and solid for TDHF results) 
and $U/J=0.0$ (dashed) with  scaling factors
 $\gamma=1.0$ (black), $\gamma=0.6$ (red),  $\gamma=0.4$ (blue), and  
$\gamma=0.2$ (blue).
}
\label{pl11.eps}
\end{figure}
 In the following
 we present pump-and-probe results for $\alpha_{\vecQ}=0.4$ where 
exact and Hartree Fock ground states show a finite charge order.
  We first consider the charge response function as it has been defined 
in the main 
 text, see Eq.~(\ref{rtz1}). The exact and TDHF results for
 $\langle \delta n_{\vecQ}  \rangle$  are displayed for $U/J=1.5$ and $U/J=0.0$ 
in Fig.~\ref{pl10.eps} as a function of the probe frequency
 $\omega$ and for several scaling factors $\gamma$ (for the TDHF results). 
We show only one exact curve for each $U$ because 
 it appears to be largely independent of the initial state at time 
 $t=0$. 
   This is different from the Hartree-Fock curves which show a shift towards 
 smaller frequencies and a significant gain in spectral weight. The 
 TDHF behavior of the two-site model  therefore resembles that of the
 macroscopic systems which we investigate in the main text. 

In Fig.~\ref{pl11.eps} we show the corresponding results for the 
superconducting response function. Again, the exact curves are independent 
 of the initial state. In this case, the TDHF shows a rather similar behavior.
 This, however, does not come as a surprise because it resembles our observations
  of the macroscopic systems in the main text.



\end{document}